%% file: arxiv16-expert-finding.tex
\pdfoutput=1
\documentclass{sig-alternate-2013}

\permission{This is an electronic version of an Article published in the 25th International World Wide Web Conference.\\\textcopyright{} 2016 International World Wide Web Conference Committee.}
\conferenceinfo{WWW 2016,}{April 11--15, 2016, Montr\'eal, Qu\'ebec, Canada.}
\copyrightetc{ACM \the\acmcopyr}
\crdata{978-1-4503-4143-1/16/04. \\
http://dx.doi.org/10.1145/2872427.2882974}

\usepackage[
  pass,
]{geometry}

\usepackage{times}
\usepackage{enumitem}
\usepackage{amsmath}
\usepackage{booktabs}
\usepackage{color}
\usepackage{eso-pic}
\usepackage{footnote}
\usepackage{grffile}
\usepackage{xifthen}
\usepackage{multirow}
\usepackage[square,comma,numbers,sort&compress,sectionbib]{natbib}
\usepackage[np, autolanguage]{numprint}
\usepackage{paralist}
\usepackage{scrpage2}
\usepackage{setspace}
\usepackage{subfig}
\usepackage{url}
\usepackage[flushleft]{threeparttable}
\usepackage[labelfont=bf,textfont=bf]{caption}
\usepackage{flushend}
\usepackage{nohyperref}

\def \paperImplementationUrl {\url{https://github.com/cvangysel/SERT}}

\def \ResearchQuestionOne {How does our discriminative log-linear model compare to vector space-based methods and generative language models for the expert retrieval task in terms of retrieval performance?}
\def \ResearchQuestionTwo {What can we learn regarding the different types of errors made by generative and discriminative language models?}
\def \ResearchQuestionThree {How does the complexity of inference in our log-linear model compare to vector-space based and generative models?}
\def \ResearchQuestionFour {How does the log-linear model handle incremental indexing and what are its limitations?}

\DeclareMathOperator*{\argmax}{arg\,max}

\newcommand{\Document}{d}
\newcommand{\EmbeddingSize}{e}

\newcommand{\VocabularyMatrix}[1][]{
\ifthenelse{\isempty{#1}}%
{W_{p}}%
{W_{p_{#1}}}%
}

\newcommand{\CandidateMatrix}[1][]{
\ifthenelse{\isempty{#1}}%
{W_{c}}%
{W_{c_{#1}}}%
}

\newcommand{\CandidateBias}[1][]{
\ifthenelse{\isempty{#1}}%
{\boldsymbol{b_c}}%
{b_{c_#1}}%
}

\newcommand{\CandidateVector}[1][]{
\ifthenelse{\isempty{#1}}%
{\boldsymbol{c}}%
{c_#1}%
}

\newcommand{\assoc}[1][]{C_{\Document{}#1}}

\newcommand{\LossFn}[1][]{L(\VocabularyMatrix{}#1, \CandidateMatrix{}#1, \boldsymbol{\CandidateBias{}}#1)}

\makeatletter
\newcolumntype{"}{@{\hskip\tabcolsep\vrule width 1pt\hskip\tabcolsep}}
\makeatother

\allowdisplaybreaks

\begin{document}

\title{Unsupervised, Efficient and Semantic Expertise Retrieval}

\numberofauthors{1}
\author{
\alignauthor
\mbox{}\hfill
\begin{tabular}{c}
    Christophe Van Gysel\\
   \email{cvangysel@uva.nl}
\end{tabular}
\hfill
\begin{tabular}{c}
    Maarten de Rijke\\
   \email{derijke@uva.nl}
\end{tabular}
\hfill
\begin{tabular}{c}
    Marcel Worring\\
   \email{m.worring@uva.nl}\\
\end{tabular}
\hfill\mbox{}\\[1.1ex]
   \affaddr{University of Amsterdam, Amsterdam, The Netherlands}
}

\maketitle
\begin{abstract}
We introduce an unsupervised discriminative model for the task of retrieving experts in online document collections. We exclusively employ textual evidence and avoid explicit feature engineering by learning distributed word representations in an unsupervised way. We compare our model to state-of-the-art unsupervised statistical vector space and probabilistic generative approaches. Our proposed log-linear model achieves the retrieval performance levels of state-of-the-art document-centric methods with the low inference cost of so-called profile-centric approaches. It yields a statistically significant improved ranking over vector space and generative models in most cases, matching the performance of supervised methods on various benchmarks. That is, by using solely text we can do as well as methods that work with external evidence and/or relevance feedback. A contrastive analysis of rankings produced by discriminative and generative approaches shows that they have complementary strengths due to the ability of the unsupervised discriminative model to perform semantic matching.
\end{abstract}






\input{introduction}
\input{related_work}
\input{methodology}
\input{experiments}

\input{discussion}
\input{conclusions}
\input{appendix}

\renewcommand{\bibsection}{
 \section*{REFERENCES}
}
\setlength{\bibsep}{0.0pt}
\bibliographystyle{abbrvnatnourl}
{\fontsize{8.5pt}{10.5pt}\selectfont
 \raggedright
 \bibliography{arxiv16-expert-finding}
}

\end{document}

%% file: introduction.tex

\section{Introduction}
\label{section:introduction}

The transition to the knowledge and information economy~\citep{OECD1996} introduces a great reliance on cognitive capabilities~\citep{Powell2004}. It is crucial for employers to facilitate information exchange and to stimulate collaboration~\citep{Davenport1998}. In the past, organizations would set-up special-purpose database systems for their members to maintain a profile \citep{BecerraFernandez2000}. However, these systems required employees to be proactive. In addition, self-assessments are known to diverge from reality \citep{Kruger1991,Berendsen2013} and document collections can quickly become practically infeasible to manage manually. Therefore, there has been an active interest in automated approaches for constructing expertise profiles~\citep{BecerraFernandez2000,TREC2010} and retrieving experts from an organization's heterogeneous document repository~\citep{Craswell2001}. \emph{Expert finding} (also known as expertise retrieval or expert search) addresses the task of finding the right person with the appropriate skills and knowledge~\citep{Balog2012}. It attempts to provide an answer to the question:
\begin{quote}
\noindent Given a topic X, who are the candidates with the most expertise w.r.t.\ X?
\end{quote}

\noindent%
The expertise retrieval task gained popularity in the research community during the TREC Enterprise Track \citep{TREC2010} and has remained relevant ever since, while broadening to social media and to tracking the dynamics of expertise~\citep{Petkova2006,Fang2007,Balog2009,Demartini2009,Fang2010,Moreira2011,Balog2012,Berendsen2013,Fang2014,vanDijk2015}. Existing methods fail to address key challenges:
\begin{inparaenum}[(1)]
	\item Queries and expert documents use different representations to describe the same concepts \citep{Hinton1986,Li2014}. Term mismatches between queries and experts \citep{Li2014} occur due to the inability of widely used maximum-likelihood language models to make use of \emph{semantic} similarities between words \citep{Salakhutdinov2009}.
	\item As the amount of available data increases, the need for more powerful approaches with greater learning capabilities than smoothed maximum-likeli\-hood language models is obvious~\citep{Vapnik1998}.
	\item Supervised methods for expertise retrieval \citep{Fang2010,Moreira2011} were introduced at the turn of the last decade. However, the acceleration of data availability has the major disadvantage that, in the case of supervised methods, manual annotation efforts need to sustain a similar order of growth. This calls for the further development of \emph{unsupervised} methods.
	\item In some expertise retrieval methods, a language model is constructed for every document in the collection. These methods lack \emph{efficient} query capabilities for large document collections, as each query term needs to be matched against every document~\citep{Balog2012}.
\end{inparaenum}
Our proposed solution has a strong emphasis on \emph{unsupervised model construction}, \emph{efficient query capabilities} and \emph{semantic matching} between query terms and candidate experts.

Specifically, we propose an unsupervised log-linear model with efficient inference capabilities for the expertise retrieval task. We show that our approach improves retrieval performance compared to vector space-based and generative language models, mainly due to its ability to perform semantic matching \citep{Li2014}. Our method does not require supervised relevance judgments and is able to learn from raw textual evidence and document-candidate associations alone. The purpose of this work is to provide insight in how discriminative language models can improve performance of core retrieval tasks compared to maximum-likelihood language models. Therefore, we avoid explicit feature engineering and the incorporation of external evidence in this paper.
In terms of performance, the current best-performing formal language model \citep{Balog2006} exhibits a worst-case time complexity linear in the size of the document collection. In contrast, the inference time complexity of our approach is asymptotically bounded by the number of candidate experts.

Our research questions are as follows:
\begin{inparaenum}[(1)]
	\item \ResearchQuestionOne{}
	\item \ResearchQuestionTwo{}
	\item \ResearchQuestionThree{}
	\item \ResearchQuestionFour{}
\end{inparaenum}

We contribute:
\begin{inparaenum}[(i)]
	\item An unsupervised log-linear model with efficient inference capabilities for the expertise retrieval task, together with an open-source implementation.\footnote{\paperImplementationUrl{}}
	\item Comparison of the retrieval performance of the log-linear model with traditional vector space-based models and language model methods on well-known benchmarks.
	\item Insights in how certainty in predictions by the log-linear model correlates with performance.
	\item In-depth analysis of the inference complexity of the log-linear model.
	\item Comparative error analysis between the semantic log-linear model and traditional generative language models that perform exact matching.
	\item Insight in the relative strengths of semantic matching and exact matching for the expert retrieval task.
\end{inparaenum}

The remainder of this paper is organized as follows. In Section~\ref{section:related}, we briefly discuss related work. Section~\ref{section:methodology} introduces the log-linear model for the expert retrieval problem. In Section~\ref{section:setup} we state our research questions and detail our experimental set-up and implementations. We provide an overview of our experimental results, followed by a discussion and analysis in Section~\ref{section:discussion}. Section~\ref{section:conclusions} concludes the paper and discusses ideas for future work.

%% file: related_work.tex

\section{Related work}
\label{section:related}

We first discuss prior work on expert retrieval and its relation to document retrieval. Then, we review semantic search methods. The ideas we present in this paper are inspired by neural language models used in automated speech recognition (ASR) and natural language processing (NLP). Consequently, we also review work from those fields.

\subsection{Expert retrieval}

Early expert retrieval systems were often referred to as expert locator and expertise management systems~\citep{Maybury2006}. These database systems often relied on people to self-assess their expertise against a predefined set of topics~\citep{McDonald2000}, which is known to generate unreliable results~\citep{BecerraFernandez2000}.

With the introduction of the P@NOPTIC system~\citep{Craswell2001}, and later the TREC Enterprise track~\citep{TREC2010}, there has been an active research interest in automated expertise profiling methods. It is useful to distinguish between \emph{profile-based} methods, which create a textual representation of a candidate's knowledge, and \emph{document-based} methods, which represent candidates as a weighted combination of documents. The latter generally performs better at ranking, while the former is more efficient as it avoids retrieving all documents relevant to a query~\citep[p.~221]{Balog2012}.

There has been much research on generative probabilistic models for expert retrieval \citep{Petkova2006,Fang2007}. Such models have been categorized in candidate generation models \citep{Cao2005}, topic generation models \citep{Balog2006,Balog2009} and proximity-based variants \citep{Serdyukov2008,Balog2009}. Of special relevance to us are the unsupervised profile-centric (Model~1) and document-centric (Model~2) models of \citet{Balog2006}, which focus on raw textual evidence without incorporating collection-specific information (e.g., query modeling, document importance or document structure). Supervised discriminative models \citep{Fang2010,Moreira2011,Sorg2011} are preferred when query-candidate relevance pairs are available for training. Unlike their generative counterparts these models have no issue combining complex and heterogeneous features (e.g., link-based features, document importance features, etc.); they resemble Learning to Rank (L2R) methods for document retrieval \citep{Balog2012,Liu2011}. However, a lack of training data may greatly hinder their applicability \citep[p.~179]{Balog2012}. Beyond unsupervised generative and supervised discriminative approaches, there are graph-based approaches based on random walks \citep{Serdyukov2008-2} and voting-based approaches based on data fusion~\citep{MacDonald2006}. \citet{Demartini2009} propose a vector space-based method for the entity ranking task; their framework extends vector spaces operating on documents to entities. See \citep{Balog2012} for a survey on the topic.

\subsection{Latent semantic models for document retrieval}

\citet{Li2014} note that the query document mismatch poses the most critical challenge in search. Semantic matching is an important attempt to remedy this problem. There has been much work on bridging the semantic gap for various different tasks \citep{Hinton1986,Demartini2009,Salakhutdinov2009,Karimzadehgan2010,Mikolov2013-2,Huang2013,Pennington2014,Kiros2014}. Expertise and document retrieval \citep[p.~224]{Balog2012} are closely related as performance of the latter can greatly impact that of the former \citep{Macdonald2008}. 

Latent Semantic Models (LSM) first became popular through the introduction of Latent Semantic Indexing (LSI) \citep{Deerwester1990}, followed by probabilistic LSI (pLSI) \citep{Hofmann1999}. LSMs based on neural networks \citep{Salakhutdinov2009,Huang2013,Shen2014} emerged in the last decade. \citeauthor{Salakhutdinov2009} \citep{Salakhutdinov2009} employ unsupervised deep auto-encoders to map documents to bit patterns using semantic addressing. \citeauthor{Huang2013} \citep{Huang2013} perform semantic matching of documents and queries by leveraging click-through data optimizing for Web document ranking. Further, deep models \citep{Burges2005,Deng2013} have been proposed to learn to rank \citep{Liu2011}.

\subsection{Neural language modeling}

Large-vocabulary neural probabilistic language models for modeling word sequence distributions have become very popular recently \citep{Bengio2003,Mnih2007,Mnih2009}. These models learn continuous-valued distributed representations for words, also known as embeddings \citep{Mnih2013,Mikolov2013,Pennington2014}, in order to fight the curse of dimensionality and increase generalization by introducing the expectation that similar word vectors signify semantically or syntactically similar words. Recurrent neural language models have shown to perform well in ASR~\citep{Mikolov2010}. \citet{Collobert2011} propose a unified neural network architecture for various NLP tasks. Even more recently, there has been a surge in multimodal neural language models \citep{Kiros2014}, which lend themselves to the task of automated image captioning.

\medskip\noindent%
What we add on top of the related work described above is the following. In this work we model the conditional probability of the expertise of a candidate given a single query term (contrary to binary relevance given a character-based n-gram \citep{Huang2013}). In the process we learn a distributed vector representation (similar to LSI, pLSI and semantic hashing) for both words and candidates such that nearby representations indicate semantically similar concepts.

We propose a log-linear model that is similar to neural language models. The important difference is that we predict a candidate expert instead of the next word. To the best of our knowledge, we are the first to propose such a solution. We employ an embedding layer in our shallow model for the same reasons as mentioned above: we learn continuous word representations that incorporate semantic and syntactic similarity tailored to an expert's domain.

%% file: methodology.tex

\break
\section{A log-linear model for\\expert search}
\label{section:methodology}

In the setting of this paper we have a document collection $D$ and a predefined set of candidate experts $C$ (entities to be retrieved). Documents $\Document{} \in D$ are represented as a sequence of words $w_1$, \ldots, $w_{|\Document{}|}$ originating from a vocabulary $V$, where $w_i \in V$ and the operator $|\cdot|$ denotes the document length in tokens. For every document $\Document{} \in D$ we write $\assoc$ to denote the set of candidates $c \in C$ associated with it (i.e., $C = \bigcup_{\Document{} \in D} \assoc{}$). These document-candidate associations can be obtained explicitly from document meta-data (e.g., the author of an e-mail) or implicitly by mining references to candidates from the document text. Notice that some documents might not be associated with any candidate. When presented with a query $q$ of constituent terms $t_1$, \ldots, $t_{|q|}$, the expert retrieval task is to return a list of candidates $\rho(C)$ ordered according to topical expertise. We generate this ranking using a relatively shallow neural network which directly models $P(\CandidateVector[] \mid q)$.

We employ vector-based distributed representations \citep{Hinton1986}, for both words (i.e., word embeddings) and candidate experts, in a way that motivates the unsupervised construction of features that express regularities of the expertise finding domain. These representations can capture the similarity between concepts (e.g., words and candidate experts) by the closeness of their representations in vector space. That is, concepts with similar feature activations are interpreted by the model as being similar, or even interchangeable.

\subsection{The model}
To address the expert search task, we model $P(\CandidateVector[j] \mid q)$ and rank candidates $c_j$ accordingly for a given $q$. We propose an unsupervised, discriminative approach to obtain these probabilities. We construct our model solely from textual evidence: we do not require query-candidate relevance assessments for training and do not consider external evidence about the corpus (e.g., different weightings for different sub-collections), the document (e.g., considering certain parts of the document more useful) nor link-based features.

Let $\EmbeddingSize{}$ denote the size of the vector-based distributed representations of both words in $V$ and candidate experts in $C$. These representations will be learned by the model using gradient descent \cite{Mikolov2013} (Section~\ref{section:learning}). For notational convenience, we write $P(\CandidateVector[] \mid \cdot)$ for the (conditional) probability distribution over candidates, which is the result of vector arithmetic. We define the probability of a candidate $c_j$ given a single word $w_i \in V$ as the log-linear model
\newcommand{\OneHotVector}[1]{\boldsymbol{v_#1}}
\newcommand{\eqUnigram}{\exp\left(\CandidateMatrix[] \cdot (\VocabularyMatrix[] \cdot \OneHotVector{i}) + \CandidateBias[]\right)}
\begin{equation}
\label{eq:unigram}
P(\CandidateVector[] \mid w_i) = \frac{1}{Z_1} \eqUnigram{},
\end{equation}
where $\VocabularyMatrix[]$ is the $\EmbeddingSize{} \times |V|$ projection matrix that maps the one-hot representation (i.e., 1-of-$|V|$) of word $w_i$, $\OneHotVector{i}$, to its $\EmbeddingSize{}$-dimensional distributed representation, $\CandidateBias[]$ is a $|C|$-dimensional bias vector and $\CandidateMatrix[]$ is the $|C| \times \EmbeddingSize{}$ matrix that maps the word embedding to an unnormalized distribution over candidates $C$, which is then normalized by $Z_1 = \sum^{|C|}_{j=1} \left[\eqUnigram\right]_j$. If we consider Bayes' theorem, the transformation matrix $\CandidateMatrix[]$ and bias vector $\CandidateBias[]$ can be interpreted as the term log-likelihood $\log P(w_i \mid \CandidateVector[])$ and candidate log-prior $\log P(\CandidateVector[])$, respectively. The projection matrix $\VocabularyMatrix[]$ attempts to soften the curse of dimensionality introduced by large vocabularies $V$ and maps words to word feature vectors \citep{Bengio2003}. Support for large vocabularies is crucial for retrieval tasks \citep{Salakhutdinov2009,Huang2013}.

We then assume conditional independence of a candidate's expertise given an observation of data (i.e., a word). Given a sequence of words $w_1$, \ldots, $w_k$ we have:
\newcommand{\eqBag}[1][]{\exp\left(\sum_{i = 1}^{k} \log\left(P(\CandidateVector[#1] \mid w_i)\right)\right)}
\newcommand{\ergkrap}{\mbox{}\hspace*{-3mm}}
\begin{eqnarray}
P(\CandidateVector[] \mid w_1, \ldots, w_k) & \ergkrap= & \ergkrap\frac{1}{Z_2} \tilde{P}(\CandidateVector[] \mid w_1, \ldots, w_k) = \frac{1}{Z_2} \prod_{i = 1}^{k} P(\CandidateVector[] \mid w_i)
\nonumber\\
& \ergkrap= &\ergkrap\frac{1}{Z_2} \eqBag[]
\label{eq:bag}
\end{eqnarray}
where $\tilde{P}(\CandidateVector[] \mid w_1, \ldots, w_k)$ denotes the unnormalized score and $Z_2 = \sum^{|C|}_{j = 1} \eqBag[j]$ is a normalizing term. The transformation to log-space in \eqref{eq:bag} is a well-known trick to prevent floating point underflow \citep[p. 445]{Montavon2012}. Given \eqref{eq:bag}, inference is straight-forward. That is, given query $q = t_1$, \ldots, $t_k$, we compute $P(\CandidateVector[] \mid t_1$, \ldots, $t_k)$ and rank the candidate experts in descending order of probability.

Eq.~\ref{eq:unigram} defines a neural network with a single hidden layer. We can add additional layers. Preliminary experiments, however, show that the shallow log-linear model \eqref{eq:unigram} performs well-enough in most cases. Only for larger data sets did we notice a marginal gain from adding an additional layer between projection matrix $\VocabularyMatrix[]$ and the softmax layer over $C$ ($\CandidateMatrix[]$ and bias $\CandidateBias[]$), at the expense of longer training times and loss of transparency.

\subsection{Parameter estimation}
\label{section:learning}
\newcommand{\CrossEntropyTarget}[1][]{
\ifthenelse{\isempty{#1}}%
{\boldsymbol{p}}%
{\boldsymbol{p}^{(#1)}}%
}
\newcommand{\CrossEntropyApprox}[1][]{
\ifthenelse{\isempty{#1}}%
{\boldsymbol{\tilde{p}}}%
{\boldsymbol{\tilde{p}}^{(#1)}}%
}
The matrices $\VocabularyMatrix[]$, $\CandidateMatrix[]$ and the vector $\CandidateBias[]$ in \eqref{eq:unigram} constitute the parameters of our model. We estimate them using error back propagation \citep{Rumelhart1986} as follows. For every document $\Document{}_j \in D$ we construct an ideal distribution over candidates $\CrossEntropyTarget[] = P(\CandidateVector[] \mid \Document{}_j)$ based on the document-candidate associations $\assoc[_j]$ such that
\[
P(\CandidateVector[] \mid d_j) =
\begin{cases}
\frac{1}{|\assoc[_j]|}, & c \in \assoc[_j] \\
0, & c \not\in \assoc[_j]
\end{cases}
\]
We continue by extracting n-grams where $n$ remains fixed during training. For every n-gram $w_1$, \ldots, $w_n$ generated from document $\Document{}$ we compute $\CrossEntropyApprox[] = P(\CandidateVector[] \mid w_1$, \ldots, $w_n)$ using \eqref{eq:bag}. During model constructing we then optimize the cross-entropy $H(\CrossEntropyTarget[], \CrossEntropyApprox[])$ (i.e., the joint probability of the training data if $|\assoc[_j]| = 1$ for all $j$) using batched gradient descent. The loss function for a single batch of $m$ instances with associated targets $(\CrossEntropyTarget[i], \CrossEntropyApprox[i])$ is as follows:
\newcommand{\Context}[1][]{
\ifthenelse{\isempty{#1}}%
{w_{1}, \ldots, w_{n}}%
{w^{(#1)}_{1}, \ldots, w^{(#1)}_{n}}%
}
\newcommand{\SourceDocument}[1]{
\Document{}^{(#1)}
}
\begin{eqnarray}
\label{eq:loss}
\lefteqn{\LossFn{}}
\nonumber \\
& = & \ergkrap\frac{1}{m} \sum_{i = 1}^m \frac{|\Document{}_{\max}|}{|\SourceDocument{i}|}  H(\CrossEntropyTarget[i], \CrossEntropyApprox[i])
\nonumber \\
&   & \ergkrap+ \frac{\lambda}{2 m} \left( \sum_{i,j} \VocabularyMatrix[i,j]^2 + \sum_{i,j} \CandidateMatrix[i,j]^2 \right) \\
& = & \ergkrap - \frac{1}{m} \sum_{i = 1}^m  \frac{|\Document{}_{\max}|}{|\SourceDocument{i}|} \sum_{j = 1}^{|C|} P(\CandidateVector[j] \mid \SourceDocument{i}) \log\left(P(\CandidateVector[j] \mid \Context[i])\right)
\nonumber \\
&   & \ergkrap+ \frac{\lambda}{2 m} \left( \sum_{i,j} \VocabularyMatrix[i,j]^2 + \sum_{i,j} \CandidateMatrix[i,j]^2 \right),
\nonumber
\end{eqnarray}
where $\SourceDocument{i}$ refers to the document from which n-gram $\Context[i]$ was extracted, $\Document{}_{\max} = \argmax_{\Document{} \in D}{|\Document{}|}$ indicates the longest document in the collection, and $\lambda$ is a weight regularization parameter. The update rule for a particular parameter $\theta$ ($\VocabularyMatrix[]$, $\CandidateMatrix[]$ or $\CandidateBias[]$) given a single batch of size $m$ is:
\newcommand{\LearningRate}{\boldsymbol{\alpha}}
\begin{equation}
\label{eq:update}
\theta^{(t+1)} = \theta^{(t)} - \LearningRate{}^{(t)} \odot \frac{\partial \LossFn[^{(t)}]}{\partial \theta},
\end{equation}
where $\LearningRate{}^{(t)}$ and $\theta^{(t)}$ denote the per-parameter learning rate and parameter $\theta$ at time $t$, respectively. The learning rate $\LearningRate{}$ consists of the same number of elements as there are parameters; in the case of a global learning rate, all elements of $\LearningRate{}$ are equal to each other.
The derivatives of the loss function \eqref{eq:loss} are given in the Appendix.

In the next section we will discuss our experimental setup, followed by an overview of our experimental results and further analysis in Section~\ref{section:discussion}.

%% file: experiments.tex

\section{Experimental Setup}
\label{section:setup}

\subsection{Research questions}
As indicated in the introduction, we seek to answer the following research questions:

\begin{description}[topsep=0pt]
\item[RQ1] \ResearchQuestionOne{}
\end{description}

\noindent%
In particular, how does the model perform when compared against vector space-based (LSI and TF-IDF) and generative approaches (profile-centric Model~1 and document-centric Model~2)?

\begin{description}[topsep=0pt]
\item[RQ2] \ResearchQuestionTwo{}
\end{description}

\noindent%
Does the best-performing generative model simply perform slightly better on the topics for which the other models perform decent as well, or do they make very different errors? If the latter holds, an ensemble of the rankings produced by both model types might exceed performance of the individual rankings.

\begin{description}[topsep=0pt]
\item[RQ3] \ResearchQuestionThree{}
\end{description}

\noindent%
The worst-case inference cost of document-centric models makes them unattractive in online settings where the set of topics is not defined beforehand and the document collection is large. Profile-centric methods are preferred in such settings as they infer from one language model per candidate expert for every topic (i.e., a pseudo-document consisting of a concatenation of all documents associated with an expert) \citep{Balog2012}. Vector space-based methods \citep{Demartini2009} have similar problems due to the curse of dimensionality \citep{Indyk1998} and consequently their inferential time complexity is likewise asymptotically bounded by the number of experts.

\begin{description}[topsep=0pt]
\item[RQ4] \ResearchQuestionFour{}
\end{description}

\subsection{Benchmarks}
\label{sec:benchmarks}

\newcommand{\specialcell}[3][c]{\begin{tabular}[#1]{@{}#2@{}}#3\end{tabular}}

\begin{table*}
\centering
\begin{threeparttable}
	\caption{An overview of the three datasets (W3C, CERC and TU) used for evaluation and analysis.\label{tbl:benchmarks}}
	\begin{tabular}{lc@{$\,$}lc@{$\,$}lc@{$\,$}l}
	\toprule
	\multicolumn{1}{c}{} & \multicolumn{2}{c}{W3C} & \multicolumn{2}{c}{CERC} & \multicolumn{2}{c}{TU} \\
	\midrule
	Number of documents & \numprint{331037}\phantom{.00} & & \numprint{370715}\phantom{.00} & & \phantom{0}\numprint{31209}\phantom{.00} &  \\
	Average document length$^a$ & \phantom{00}\numprint{1237.23} & & \phantom{000,}\numprint{460.48} & & \phantom{00}\numprint{2454.93} & \\

	\\

	Number of candidates$^b$ & \phantom{000,}\numprint{715}\phantom{.00} & & \phantom{00}\numprint{3479}\phantom{.00} & & \phantom{000,}\numprint{977}\phantom{.00} & \\

	\\

	\specialcell{l}{Number of document-candidate associations} & \numprint{200939}\phantom{.00} & & \numprint{236958}\phantom{.00} & & \phantom{0}\numprint{36566}\phantom{.00} & \\
	\specialcell{l}{Number of documents (with $\assoc > 0$)} & \phantom{0}\numprint{93826}\phantom{.00} & & \numprint{123934}\phantom{.00} & & \phantom{0}\numprint{27834}\phantom{.00} &  \\
	\specialcell{l}{Number of associations per document$^c$} & \phantom{000,00}\numprint{2.14} & $\pm\, 3.29$ & \phantom{000,00}\numprint{1.91} & $\pm\, 3.70$ & \phantom{000,00}\numprint{1.13} & $\pm\, 0.39$ \\
	\specialcell{l}{Number of associations per candidate} & \phantom{000,}\numprint{281.03} & $\pm\, 666.63$ & \phantom{000,0}\numprint{68.11} & $\pm\,$\numprint{1120.74} & \phantom{000,0}\numprint{37.43} & $\pm\, 61.00$ \\

	\\

	Queries & \phantom{000,0}\numprint{49}\phantom{.00} & (2005) & \phantom{000,0}\numprint{50}\phantom{.00} & (2007) & \phantom{00}\numprint{1662}\phantom{.00} & (GT1) \\
	& \phantom{000,0}\numprint{50}\phantom{.00} & (2006) & \phantom{000,0}\numprint{77}\phantom{.00} & (2008) & \phantom{00}\numprint{1266}\phantom{.00} & (GT5) \\
	\bottomrule
	\end{tabular}
	\begin{tablenotes}
	\footnotesize
	\item[$a$] Measured in number of tokens.
	\item[$b$] Only candidates with at least a single document association are considered.
	\item[$c$] Only documents with at least one association are considered.
	\end{tablenotes}
\end{threeparttable}
\vspace*{-.5\baselineskip}
\end{table*}%

The proposed method is applicable in the setting of the Expert Search task of the TREC Enterprise track from 2005 to 2008 \citep{TREC2010}. We therefore evaluate on the W3C and CERC benchmarks released by the track. The W3C dataset \citep{Craswell2005} is a crawl of the W3C's sites in June 2004 (mailing lists, web pages, etc.). The CSIRO Enterprise Research Collection (CERC) \citep{Bailey2007} is a dump of the intranet of Australia's national science agency. Additionally, we evaluate our method on a smaller, more recent benchmark based on the employee database of Tilburg University (TU) \citep{Berendsen2013}, which consists of bi-lingual, heterogeneous documents. See Table~\ref{tbl:benchmarks}.

Mining document-candidate associations and how they influence performance has been extensively covered in previous work \citep{Balog2006,Balog2012} and is beyond the scope of this work. For TU, the associations are part of the benchmark. For W3C, a list of possible candidates is given and we extract the associations ourselves by performing a case-insensitive match of full name or e-mail address \citep{Balog2006}. For CERC, we make use of publicly released associations~\citep{Balog2008}.

As evaluation measures we use Mean Average Precision (MAP), Mean Reciprocal Rank (MRR), Normalized Discounted Cumulative Gain at rank 100 (NDCG@100) and Precision at rank 5 (P@5) and rank 10 (P@10).

\subsection{Baselines}
\label{sec:baseline}

We compare our approach to existing unsupervised methods for expert retrieval that solely rely on textual evidence and static doc\-ument-candidate associations.
\begin{inparaenum}[(1)]%
\item \citet{Demartini2009} propose a generic framework to adapt vector spaces operating on documents to entities. We compare our method to TF-IDF (raw frequency and inverse document frequency) and LSI (300 latent topics) variants of their vector space model for entity ranking (using cosine similarity).
\item In terms of language modeling, \citet{Balog2006} propose two models for expert finding based on generative language models. The first takes a profile-centric approach comparing the language model of every expert to the query, while the second is document-centric. We consider both models with different smoothing configurations: Jelinek-Mercer (jm) smoothing with $\lambda=0.5$ \citep{Balog2006} and Dirichlet (d) smoothing with $\beta$ equal to the average document length \citep{Balog2009} (see Table~\ref{tbl:benchmarks}).
\end{inparaenum}
Significance of results produced by the baselines (compared to our method) is determined using a two-tailed paired randomization test \citep{Smucker2007}.

\subsection{Implementation details}
\label{sec:implementation}

\newcommand{\m}{\sqrt{\frac{6.0}{m + n}}}


The vocabulary $V$ is constructed from each corpus by ignoring punctuation, stop words and case; numbers are replaced by a numerical placeholder token. During our experiments we prune $V$ by only retaining the $2^{16}$ most-frequent words so that each word can be encoded by a 16-bit unsigned integer. Incomplete n-gram instances are padded by a special-purpose token.

In terms of parameter initialization, we sample the initial matrices $\CandidateMatrix{}$ and $\VocabularyMatrix{}$ \eqref{eq:unigram} uniformly in the range
\begin{equation*}
\left[ -\m, \m \; \right]
\end{equation*}
for an $m \times n$ matrix, as this initialization scheme improves model training convergence \citep{Glorot2010}, and take the bias vector $\CandidateBias{}$ to be null. The projection layer $\VocabularyMatrix{}$ is initialized with pre-trained word representations trained on Google News data \citep{Mikolov2013-2}; the number of word features is set to $\EmbeddingSize{}=300$, similar to pre-trained representations.

We used adadelta ($\rho=0.95$, $\epsilon=10^{-6}$) \citep{Zeiler2012} with batched gradient descent ($m=1024$) and weight decay $\lambda=0.01$ during training on NVidia GTX480 and NVidia Tesla K20 GPUs. We only iterate once over the entire training set for each experiment.

%% file: discussion.tex

\section{Results and Discussion}
\label{section:discussion}

We start by giving a high-level overview of our experimental results and then address issues of scalability, provide an error analysis and discuss the issue of incremental indexing.

\subsection{Overview of experimental results}

\newcommand{\Significant}{^{*}}
\newcommand{\MoreSignificant}{^{**}}
\newcommand{\HighlySignificant}{^{***}}

\begin{table*}[t]
	\centering
	\caption{Evaluation results for models trained on the W3C, CERC and TU benchmarks. Suffixes (d) and (jm) denote Dirichlet and Jelinek-Mercer smoothing,  respectively (Section~\ref{sec:baseline}). Significance of results is determined using a two-tailed paired randomization test \citep{Smucker2007} ($\HighlySignificant{} \, p < 0.01$;  $\MoreSignificant{} \, p < 0.05$; $\Significant{} \, p < 0.1$) with respect to the log-linear model (adjusted using the Benjamini-Hochberg procedure for multiple testing \citep{Benjamini1995}).\label{tbl:all_results}}
	\resizebox{\textwidth}{!}{\input{resources/all_results.tex}}
	\vspace*{-.5\baselineskip}
\end{table*}

\begin{figure*}[t]
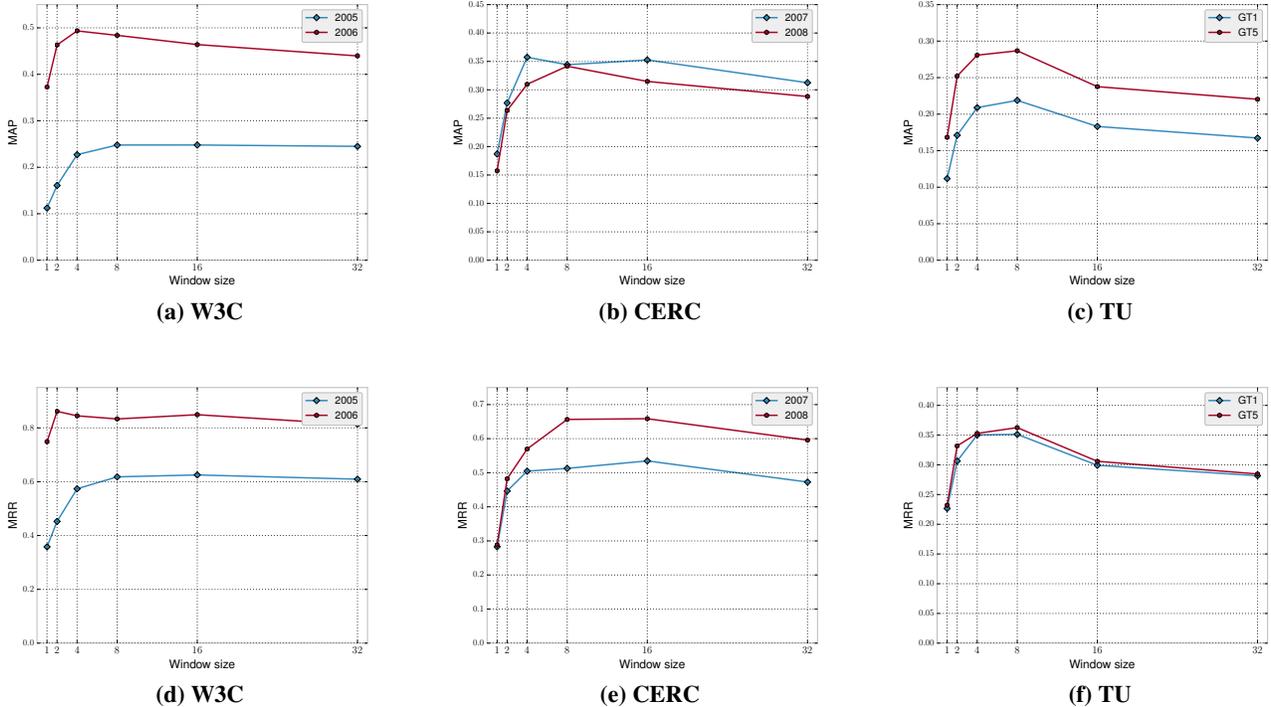

	\centering

	\newcommand{\windowsweepplot}[4]{
		\subfloat[#3\label{fig:windowsweep:#1}]{%
			\includegraphics[width=0.32\textwidth]{resources/window_sweep_results/#4_#2.pdf}
		}
	}

	\windowsweepplot{w3c}{map}{W3C}{W3C}
	\hfill
	\windowsweepplot{cerc}{map}{CERC}{CERC}
	\hfill
	\windowsweepplot{tu}{map}{TU}{TU}

	\windowsweepplot{w3c}{mrr}{W3C}{W3C}
	\hfill
	\windowsweepplot{cerc}{mrr}{CERC}{CERC}
	\hfill
	\windowsweepplot{tu}{mrr}{TU}{TU}

	\caption{Sensitivity analysis for window size (n-gram) during parameter estimation \eqref{eq:loss} for W3C, CERC and TU benchmarks.\label{fig:windowsweep}}
\vspace*{-.5\baselineskip}
\end{figure*}

We evaluate the log-linear model on the W3C, CERC and TU benchmarks (Section~\ref{sec:benchmarks}). During training we extract non-overlap\-ping n-grams for the W3C and CERC benchmarks and overlapping n-grams for the TU benchmark. As the TU benchmark is considerably smaller, we opted to use overlapping n-grams to counter data sparsity. The architecture of each benchmark model (e.g., number of candidate experts) is inherently specified by the benchmarks themselves (see Table~\ref{tbl:benchmarks}). However, the choice of n-gram size during training remains open. Errors for input $w_1$, \ldots, $w_n$ are propagated back through $\CandidateMatrix{}$ until the projection matrix $\VocabularyMatrix{}$ is reached; if a single word $w_i$ causes a large prediction error, then this will influence its neighboring words $w_1$, \ldots, $w_{i-1}$, $w_{i+1}$, \ldots, $w_{n}$ as well. This allows the model to learn continuous word representations tailored to the expert retrieval task and the benchmark domain.

A larger window size has a negative impact on batch throughput during training. We are thus presented with the classic trade-off between model performance and construction efficiency. Notice, however, that the number of n-grams decreases as the window size increases if we extract non-overlapping instances. Therefore, larger values of $n$ lead to faster wall-clock time model construction for the W3C and CERC benchmarks in our experiments.

We sweep over the window width $n = 2^i$ ($0 \leq i < 6$) for all three benchmarks and their corresponding relevance assessments. We report MAP and MRR for every configuration (see Figure~\ref{fig:windowsweep}). We observe a significant performance increase between $n = 1$ and $n = 2$ on all benchmarks, which underlines the importance of the window size parameter. The increase in MAP implies that the performance achieved is not solely due to initialization with pre-trained representations (Section~\ref{sec:implementation}), but that the model efficiently learns word representations tailored to the problem domain. The highest MAP scores are attained for relatively low $n$. As $n$ increases beyond $n = 8$ a gradual decrease in MAP is observed on all benchmarks. In our remaining experiments we choose $n = 8$ regardless of the benchmark.

\begin{figure*}[t]
	\centering

	\newcommand{\entropyprecisionplot}[5]{
		\subfloat[#2 ($R=#4\HighlySignificant{}$)\label{fig:entropyprecision:#1}]{%
			\includegraphics[width=0.32\textwidth]{resources/entropy_precision_results/#3_entropy_ap.pdf}
		}
	}

	\entropyprecisionplot{w3c}{W3C}{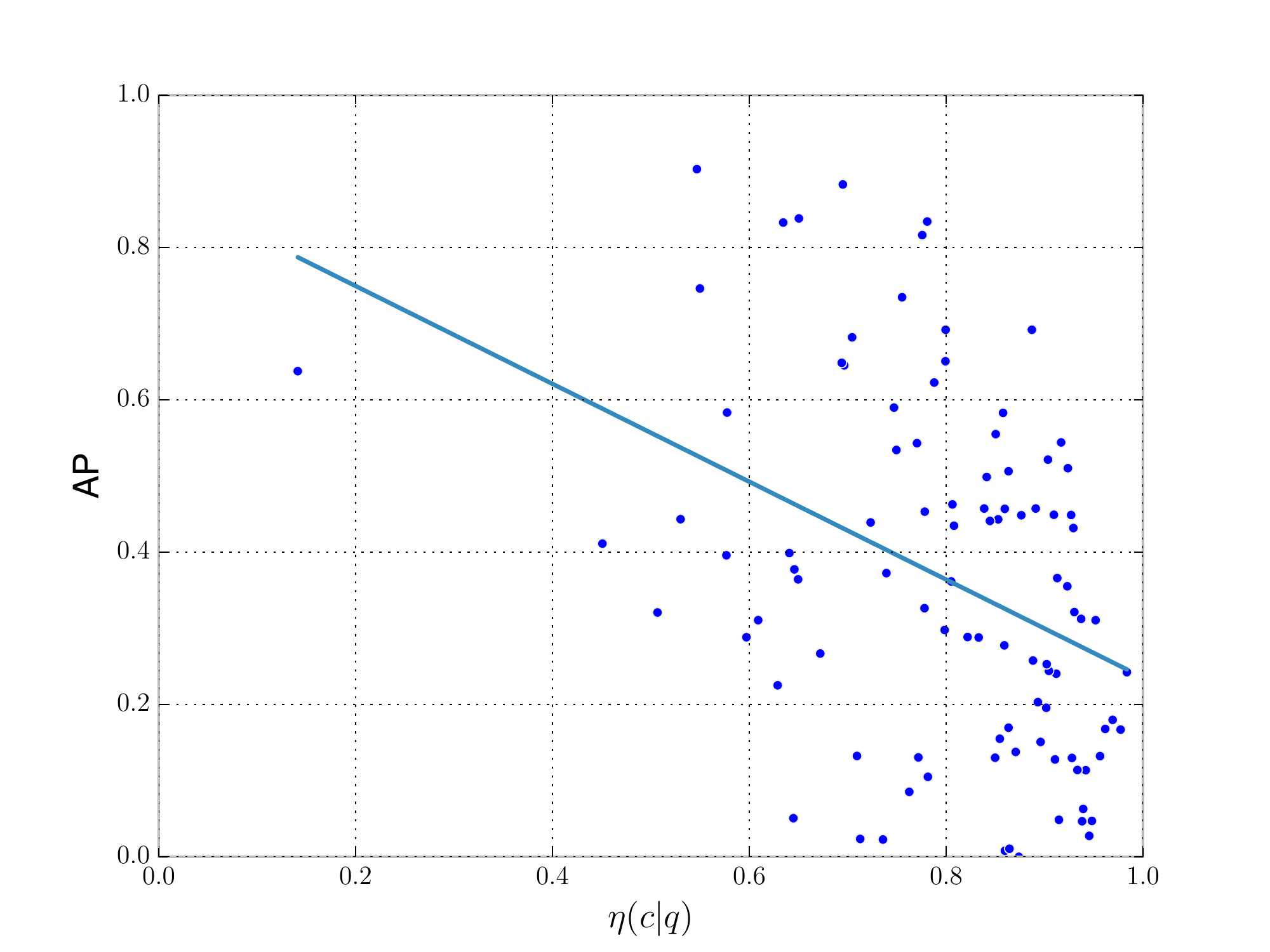}{-0.39}{0.000065}
	\hfill
	\entropyprecisionplot{cerc}{CERC}{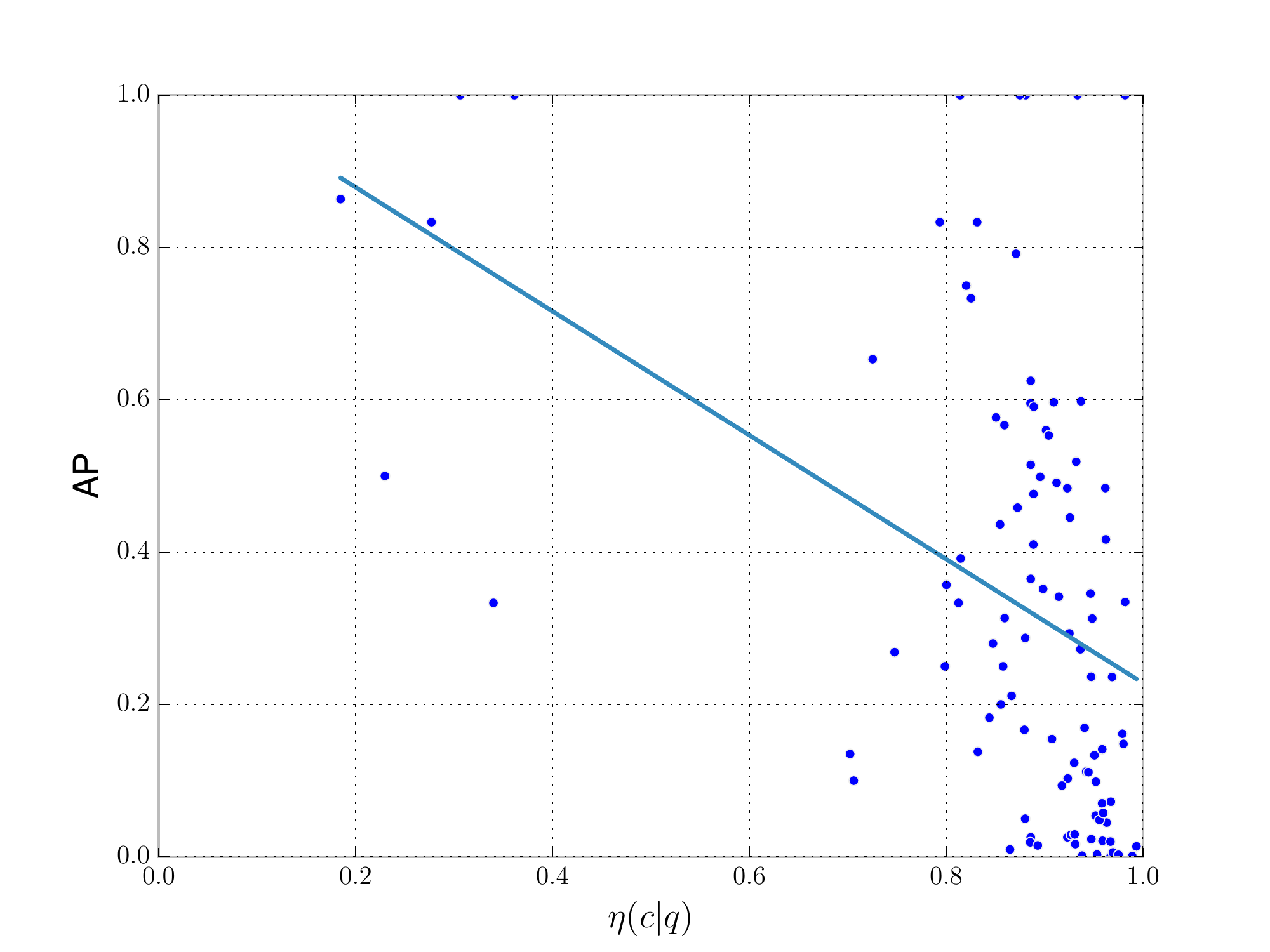}{-0.44}{0.000020}
	\hfill
	\entropyprecisionplot{tu}{TU}{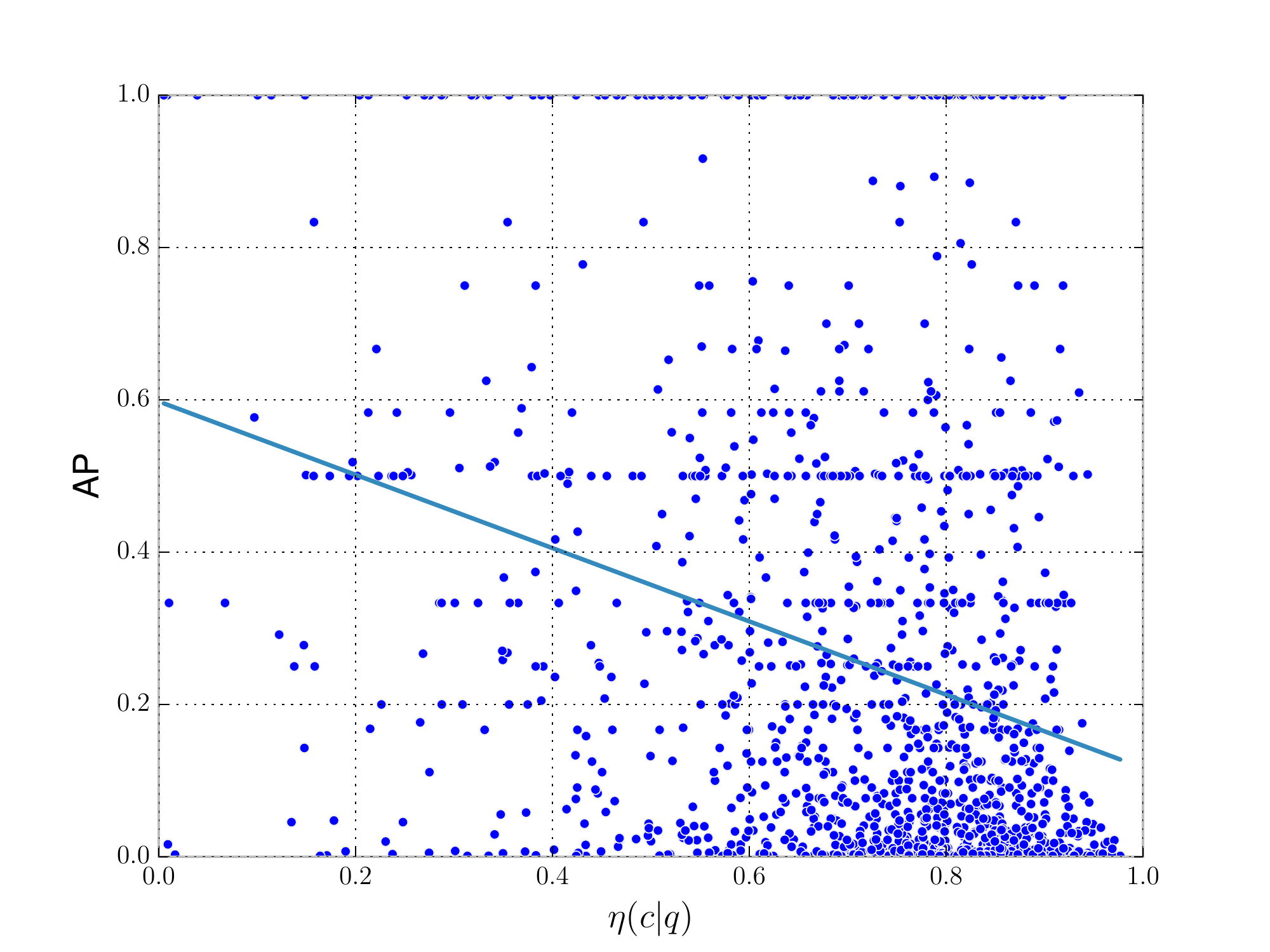}{-0.30}{0.000000}

\caption{Scatter plot of the normalized entropy of distribution $P(\CandidateVector[] \mid q)$ \eqref{eq:bag} returned by the log-linear model and per-query average precision for W3C, CERC and TU benchmarks. Pearson's $R$ and associated $p$-value (two-tailed paired permutation test: $\HighlySignificant{} \, p < 0.01; \, \MoreSignificant{} \, p < 0.05; \, \Significant{} \, p < 0.1$) are between parentheses. The depicted linear fit was obtained using an ordinary least squares regression. \label{fig:entropyprecision}}
\end{figure*}

Words $w_i$ that mainly occur in documents associated with a particular expert are learned to produce distributions $P(\CandidateVector[] \mid w_i)$ with less uncertainty than words associated with many experts in \eqref{eq:unigram}. The product of $P(\CandidateVector[] \mid w_i)$ in \eqref{eq:bag} aggregates this expertise evidence generated by query terms. Hence, queries with strong evidence for a particular expert should be more predictable than very generic queries. To quantify uncertainty we measure the normalized entropy \citep{Shannon1948} of $P(\CandidateVector[] \mid  q)$:
\begin{equation}
\label{eq:normentropy}
\eta(c \mid  q) = - \frac{1}{\log(|C|)} \sum_{j=1}^{|C|} p(\CandidateVector[j] \mid  q) \log (p(\CandidateVector[j] \mid  q)).
\end{equation}
Equation~\ref{eq:normentropy} can be interpreted as a similarity measure between the given distribution and the uniform distribution. Importantly, Figure~\ref{fig:entropyprecision} shows that there is a statistically significant negative correlation between query-wise normalized entropy and average precision for all benchmarks.

Table~\ref{tbl:all_results} presents a comparison between the log-linear model and the various baselines (Section~\ref{sec:baseline}). Our unsupervised method significantly ($p < 0.01$) outperforms the LSI-based method consistently. In the case of the TF-IDF method and the profile-centric generative language models (Model~1), we always perform better and statistical significance is achieved in the majority of cases. The document-centric language models (Model~2) perform slightly better than our method on two (out of six) benchmarks in terms of MAP and NDCG@$100$:%
\begin{inparaenum}[(1)]%
\item For the CERC 2007 assessment we match performance of the document-centric generative model with Jelinek-Mercer smoothing.
\item For TU GT1 the generative counterpart seems to outperform our method.
\end{inparaenum}

Notice that over all assessments, the log-linear model consistently outperforms all profile-centric approaches and is only challenged by the smoothed document-centric approach. In addition, for the precision-based measures (P@$k$ and MRR), the log-linear model consistently outperforms all other methods we compare to.

Next, we turn to a topic-wise comparative analysis of discriminative and generative models. After that, we analyze the scalability and efficiency of the log-linear model and compare it to that of the generative counterparts, and address incremental indexing.

\subsection{Error analysis}
\label{sec:erroranalysis}

\begin{figure*}[t]
	\centering

	\newcommand{\errorplot}[4]{
		\subfloat[#3\label{fig:error:#1}]{%
			\includegraphics[width=0.32\textwidth]{resources/error_analysis_results/#4.pdf}
		}
	}

	\errorplot{w3c-2005}{map}{W3C 2005}{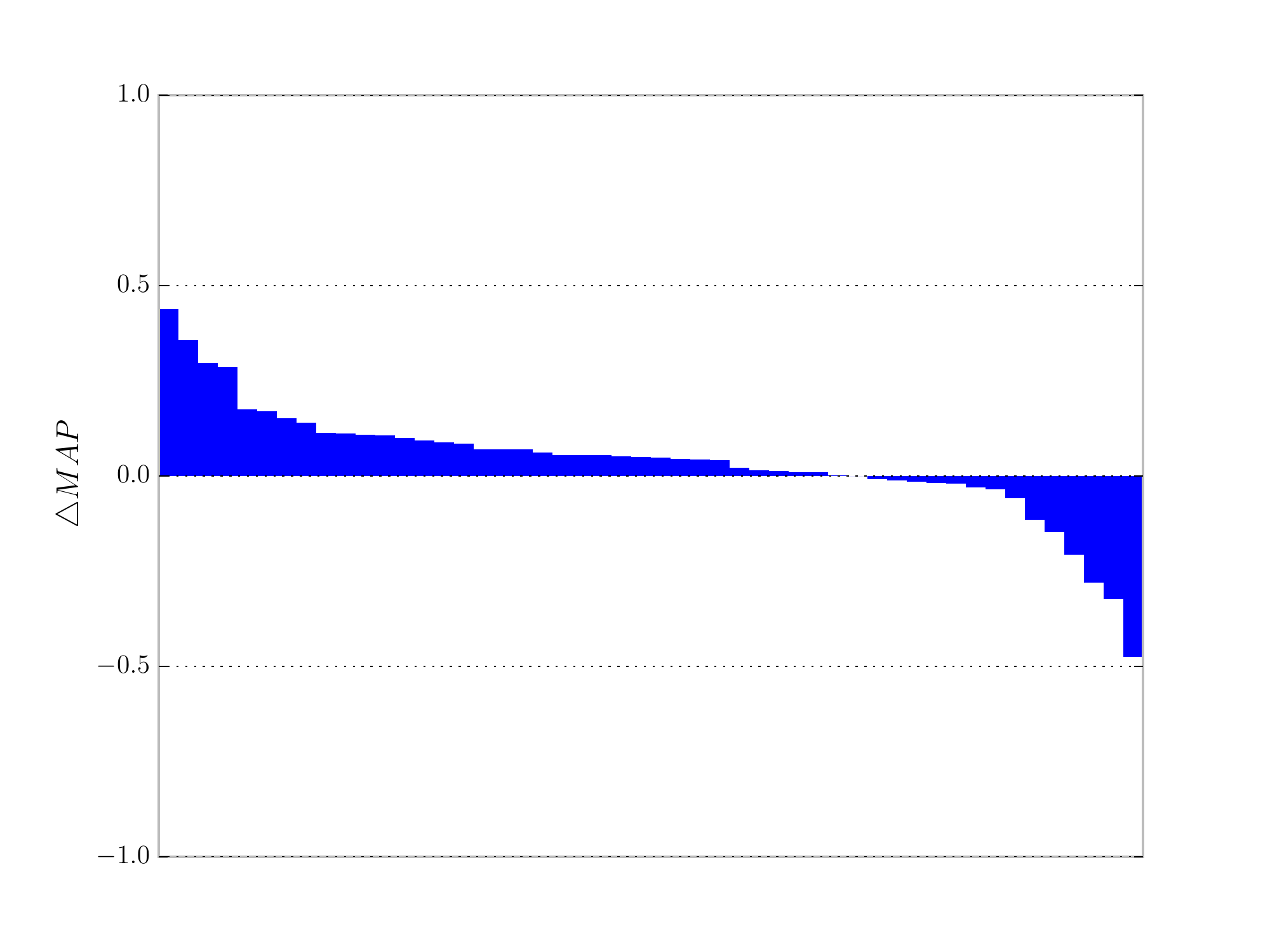}
	\hfill
	\errorplot{cerc-2007}{map}{CERC 2007}{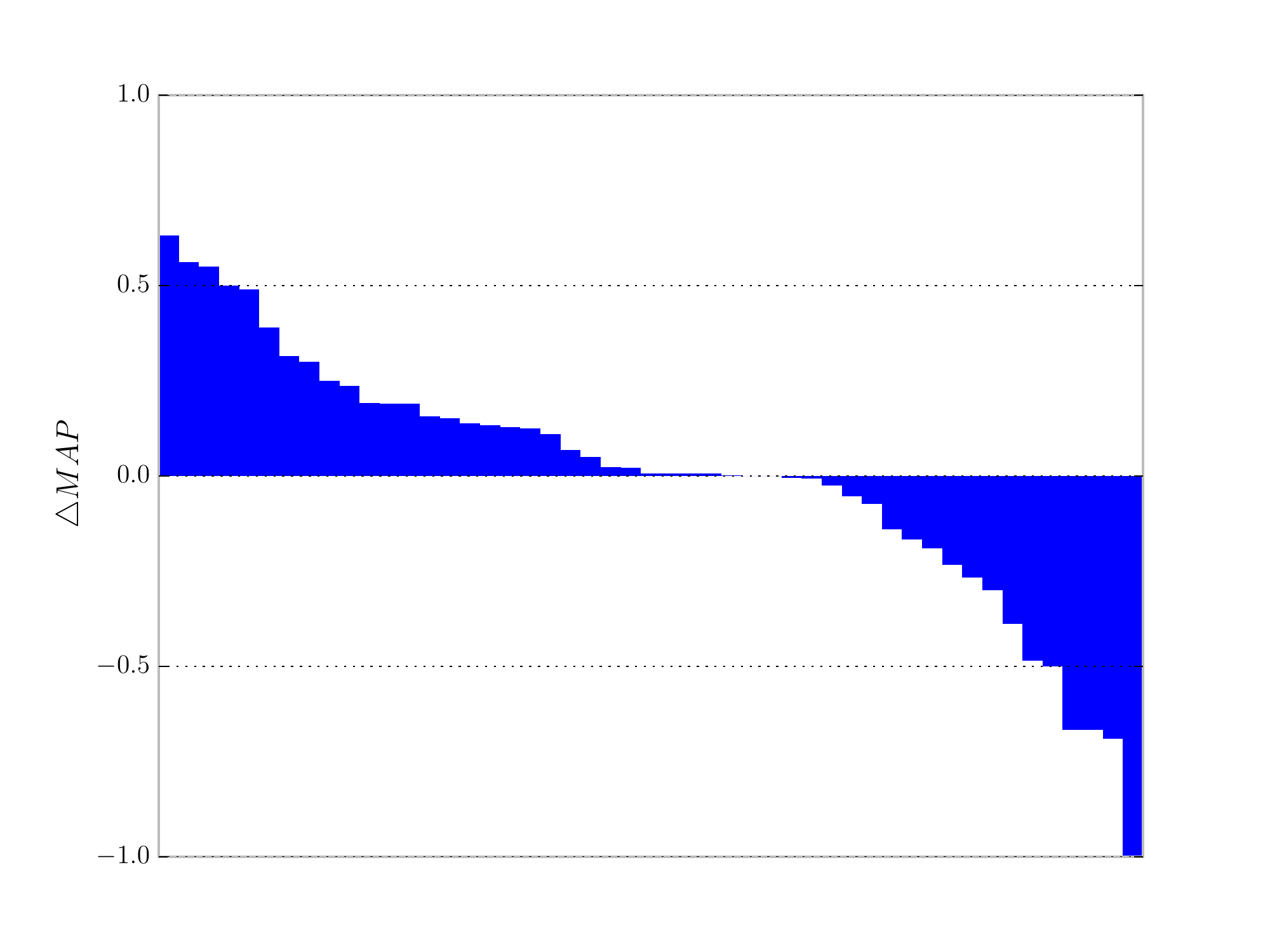}
	\hfill
	\errorplot{tu-gt1}{map}{TU GT1}{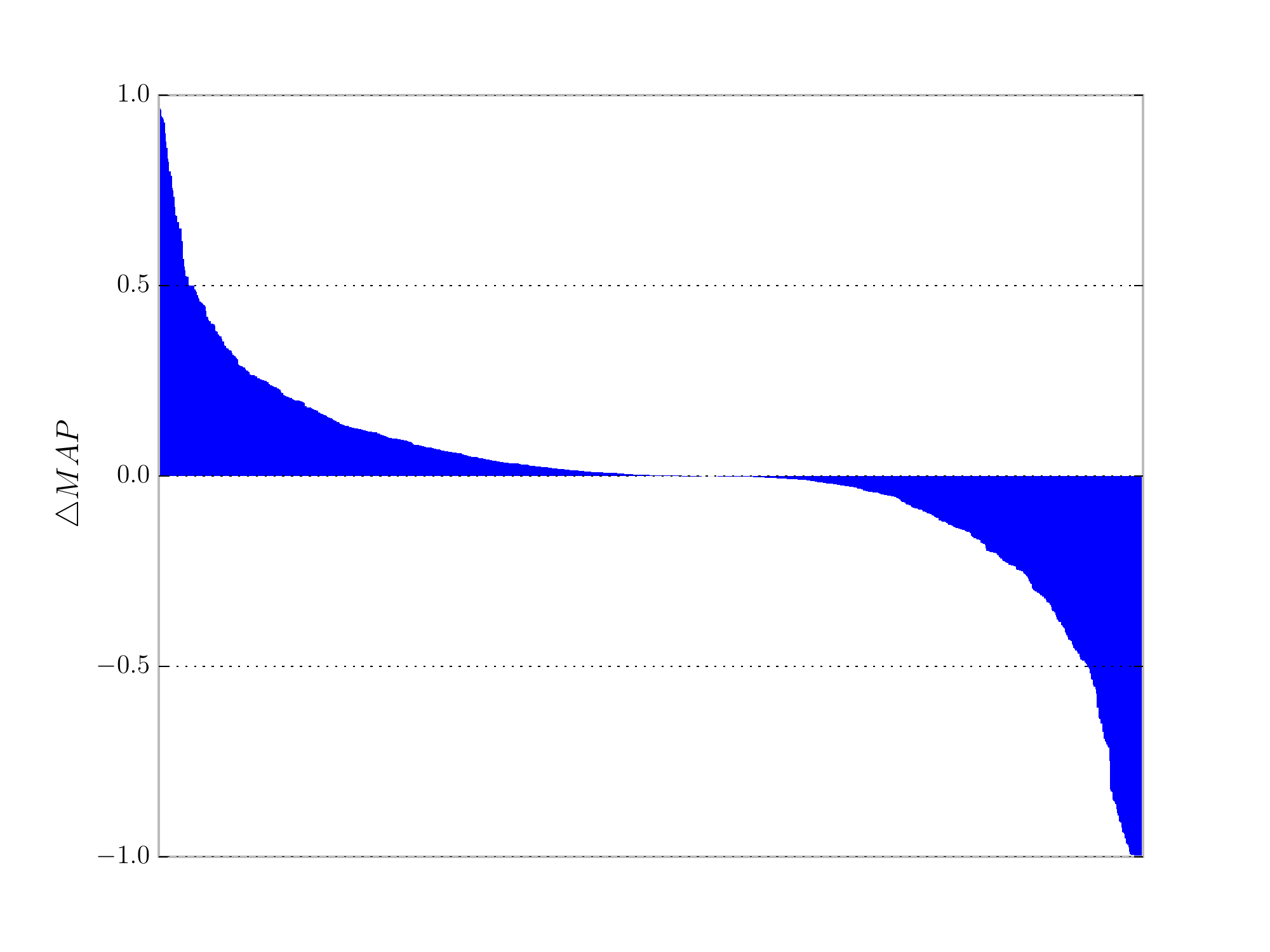}

\vspace*{-\baselineskip}
	\errorplot{w3c-2006}{map}{W3C 2006}{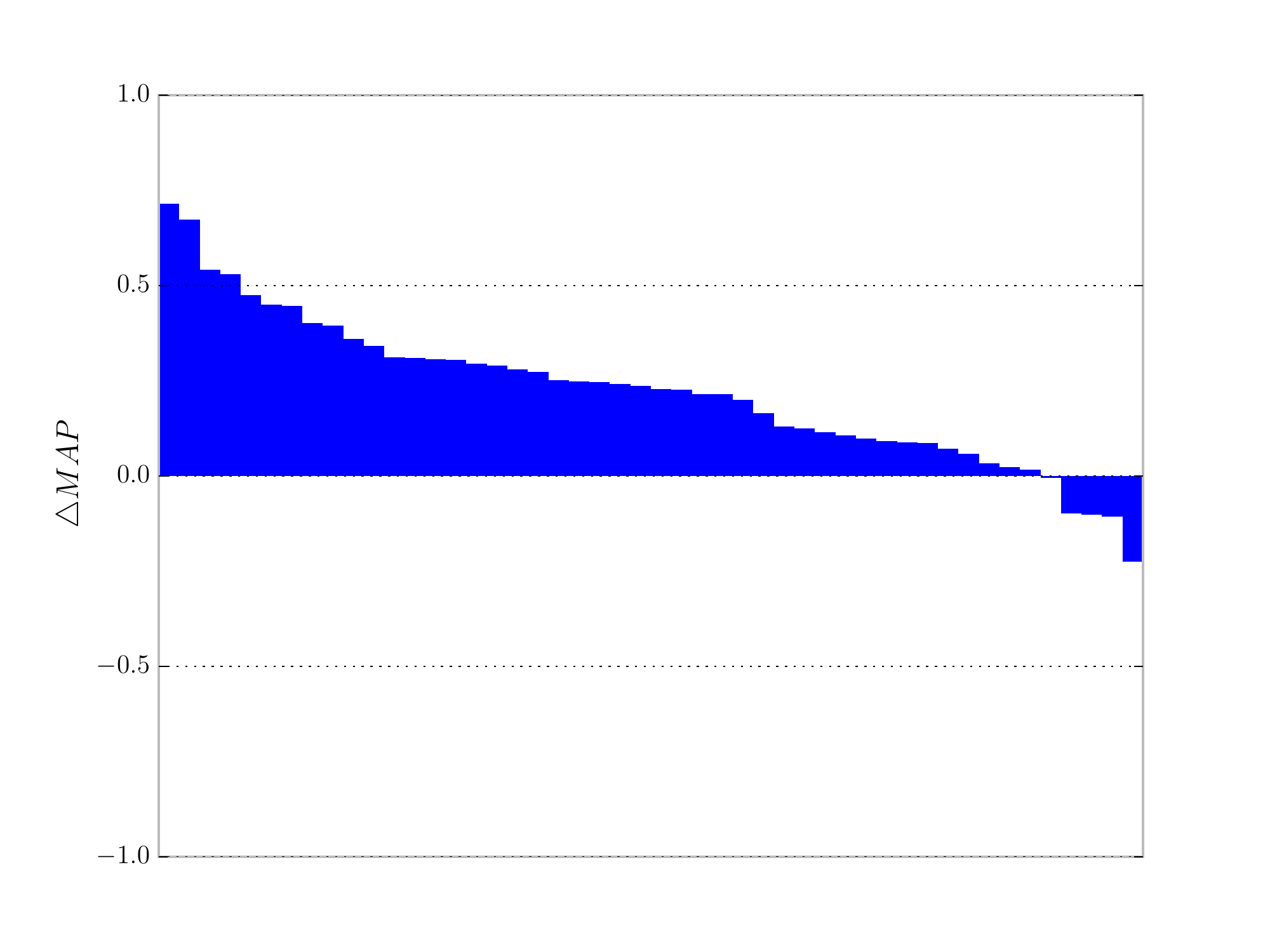}
	\hfill
	\errorplot{cerc-2008}{map}{CERC 2008}{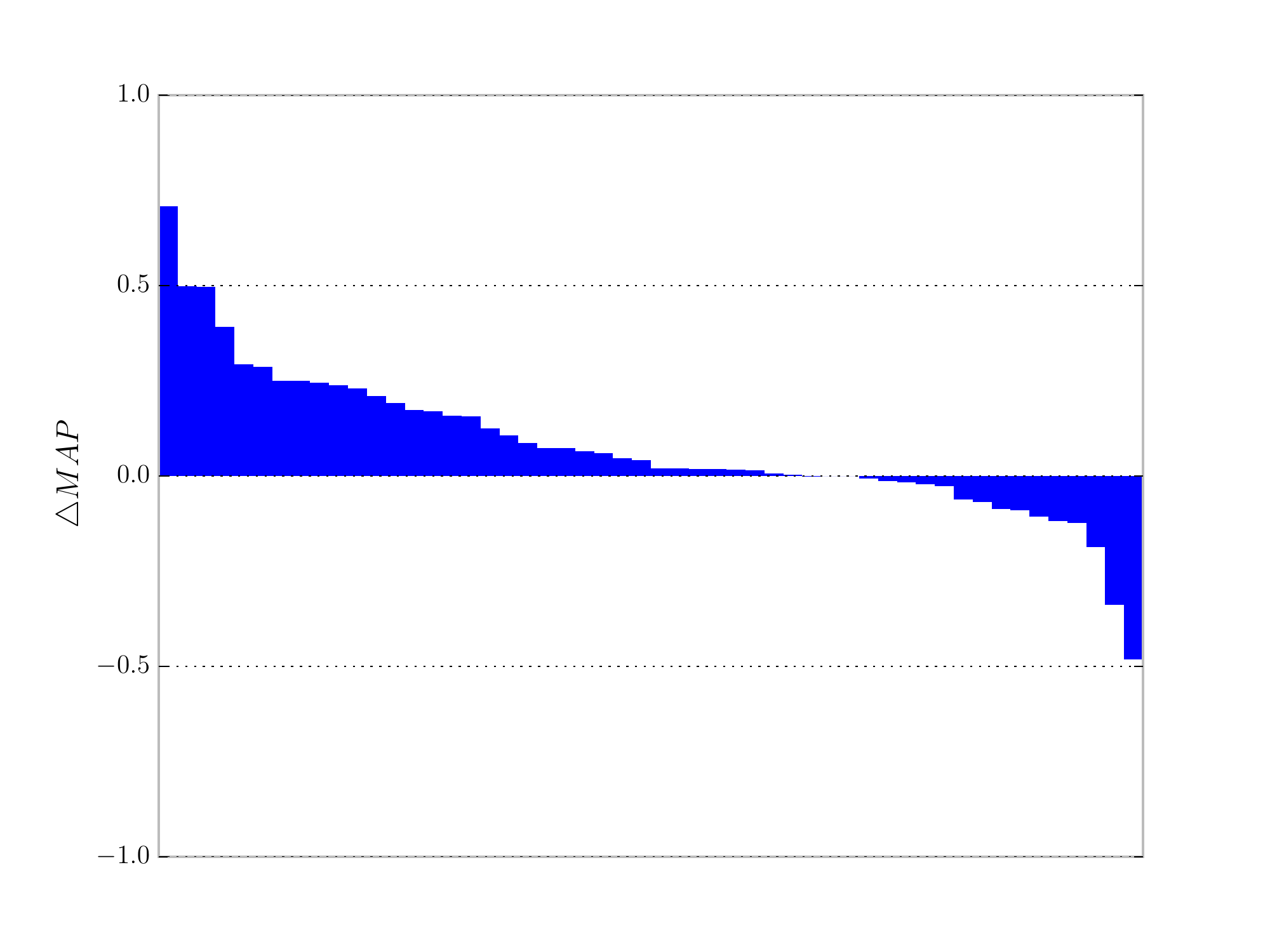}
	\hfill
	\errorplot{tu-gt5}{map}{TU GT5}{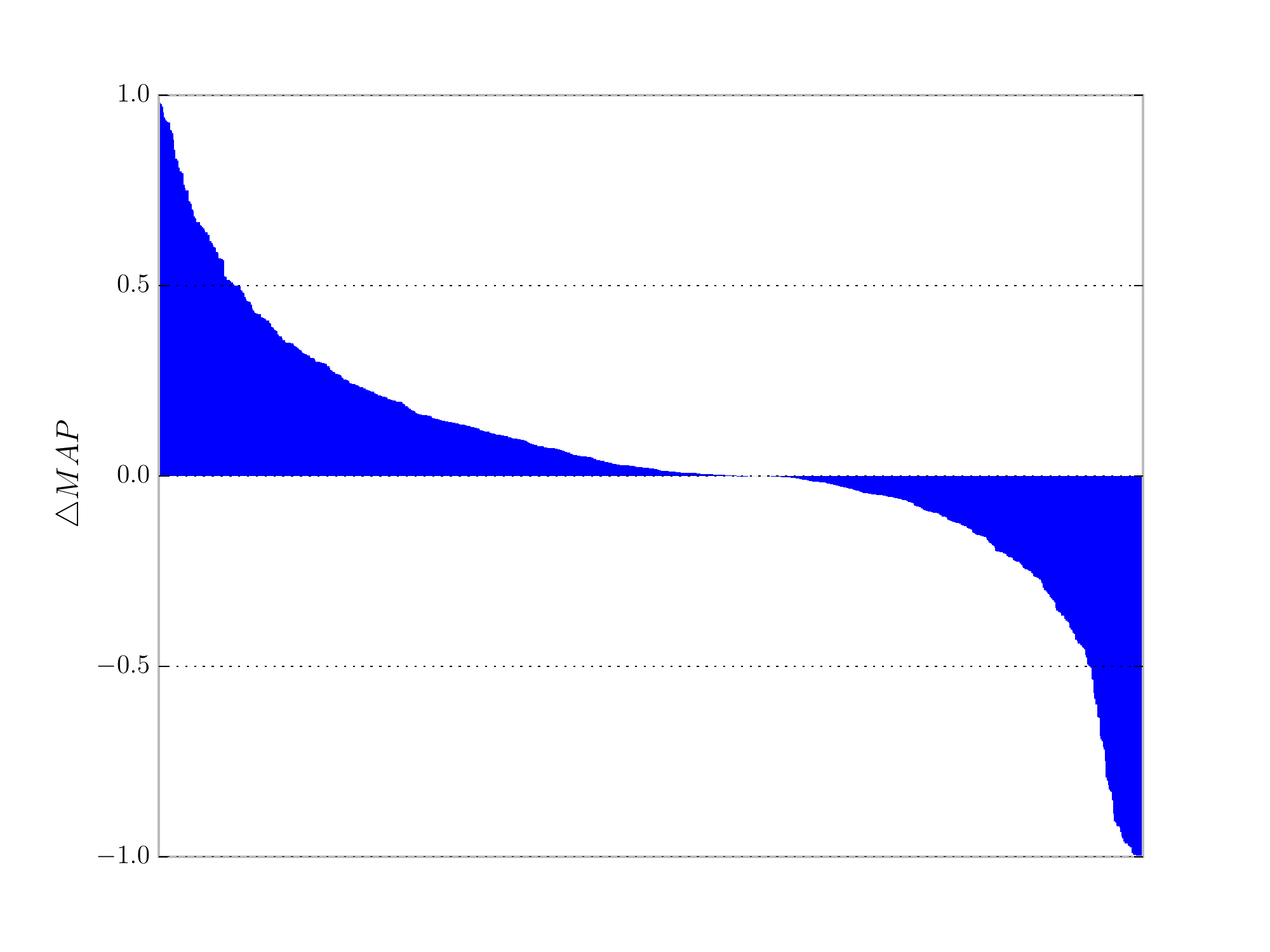}

	\caption{Difference of average precision between log-linear model and Model 2 \citep{Balog2006} with Jelinek-Mercer smoothing per topic for W3C, CERC and TU benchmarks.\label{fig:error}}
	 \vspace*{-.5\baselineskip}
\end{figure*}

How does our log-linear model achieve its superior performance over established generative models? Figure~\ref{fig:error} depicts the per-topic differences in average precision between the log-linear model and Model~2 (with Jelinek-Mercer smoothing) on all benchmarks. For each plot, the vertical bars with a positive AP difference correspond to test topics for which the log-linear model outperforms Model~2 and vice versa for bars with a negative AP difference.

The benefit gained from the projection matrix $\VocabularyMatrix{}$ is two-fold. First, it avoids the curse of dimensionality introduced by large vocabularies. Second, term similarity with respect to the expertise domain is encoded in latent word features. When examining words nearby query terms in the embedding space, we found words to be related to the query term. For example, word vector representations of \emph{xml} and \emph{nonterminal} are very similar for the W3C benchmark ($l_2$ norm). This can be further observed in Figure~\ref{fig:windowsweep}: log-linear models trained on single words perform significantly worse compared to those that are able to learn from neighboring words.


We now take a closer look at the topics for which the log-linear model outperforms Model~2 and vice versa. More specifically, we investigate textual evidence related to a topic and whether it is considered relevant by the benchmark. For the log-linear model, we examine terms nearby topic terms in $\VocabularyMatrix[]$ ($l_2$-norm), as these terms are considered semantically similar by the model and provide a means for semantic matching. For every benchmark, we first consider topics where exact matches (Model~2) perform best, followed by examples which benefit from semantic matching (log-linear model). Topic identifiers are between parentheses.

\begin{description}[itemsep=0pt,topsep=0pt]
\item[W3C] Topics \emph{P3P specification} and \emph{CSS3} (EX8 and EX69, respectively) should return candidates associated with the definition of these standards. The log-linear model, however, considers these close to related technologies such as CSS2 for \emph{CSS3} and UTF-8 for P3P. Semantic matching works for topics \emph{Semantic Web Coordination} and \emph{Annotea server protocol} (EX1 and EX103), where the former is associated with RDF libraries, RDF-related jargon and the names of researchers in the field, while the latter is associated with implementations of the protocol and the maintainer of the project.
\item[CERC] For CSIRO, topic \emph{nanohouse} (CE-035) is mentioned in many irrelevant contexts (i.e., spam) and therefore semantic matching fails. The term \emph{fish oil} (CE-126) is quickly associated with different kinds of fish, oils and organizations related to marines and fisheries. On the other hand, we observe \emph{sensor networks} (CE-018) to be associated with sensor/networking jargon and sensor platforms. Topic \emph{forensic science workshop} (CE-103) expands to syntactically-similar terms (e.g., plural), the names of science laboratories and references to support/law-protection organizations.
\break
\item[TU] The TU benchmark contains both English and Dutch textual evidence. Topics \emph{sustainable tourism} and \emph{interpolation} (1411 and 4882) do not benefit from semantic matching due to a semantic gap: \emph{interpolation} is associated with the polynomial kind while the relevance assessments focus on stochastic methods. Interestingly, for the topic \emph{law and informatization/computerization} (1719) we see that the Dutch translation of \emph{law} is very closely related. Similar terms to \emph{informatization} are, according to the log-linear model, Dutch words related to cryptography. Similar dynamics are at work for \emph{legal-political space} (12603), where translated terms and semantic-syntactic relations aid performance.
\end{description}

\begin{figure*}[t]
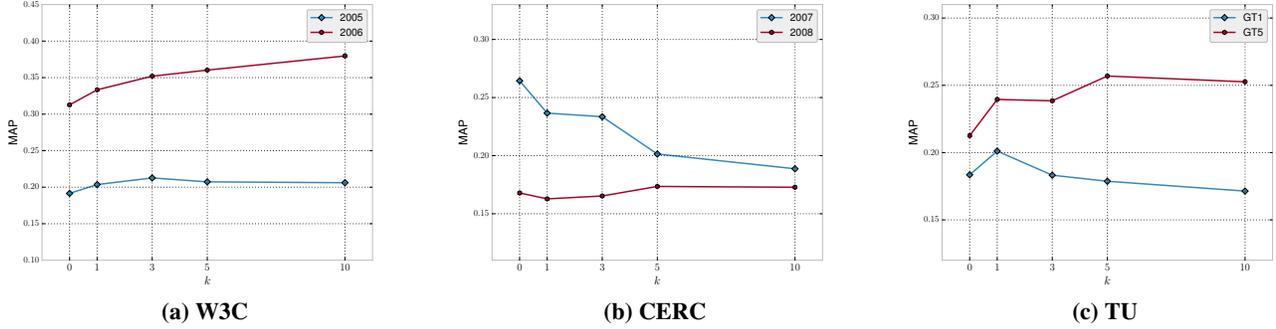

	\centering

	\newcommand{\queryexpansionplot}[4]{
		\subfloat[#3\label{fig:queryexpansion:#1}]{%
			\includegraphics[width=0.32\textwidth]{resources/semantic_expansion/#4_#2.pdf}
		}
	}

	\queryexpansionplot{w3c}{map}{W3C}{W3C}
	\hfill
	\queryexpansionplot{cerc}{map}{CERC}{CERC}
	\hfill
	\queryexpansionplot{tu}{map}{TU}{TU}

	\caption{Effect of query expansion by adding nearby terms in $\VocabularyMatrix{}$ \eqref{eq:unigram} in traditional language models (Model 1 \citep{Balog2006} with Jelinek-Mercer smoothing) for W3C, CERC and TU benchmarks.\label{fig:queryexpansion}}

\end{figure*}

\noindent%
In order to further quantify the effect of the embedding matrix $\VocabularyMatrix{}$, we artificially expand benchmark topic terms by $k$ nearby terms. We then examine how the performance of a profile-centric generative language model \citep[Model~1]{Balog2006} evolves for different values of $k$ (Figure~\ref{fig:queryexpansion}). The purpose of this analysis is to provide further insight in the differences between maximum-likelihood language models and the log-linear model. Figure~\ref{fig:queryexpansion} shows that, for most benchmarks, MAP increases as $k$ goes up. Interestingly enough, the two benchmarks that exhibit a decrease in MAP for larger $k$ (CERC 2007 and TU GT1) are likewise those for which generative language models outperform the log-linear model in Table~\ref{tbl:all_results}. This suggests that the CERC 2007 and TU GT1 benchmarks require exact term matching, while the remaining four benchmarks benefit greatly from the semantic matching provided by our model.


\begin{table*}[t]
	\centering
	\caption{Comparison of Model~2, the log-linear model and an ensemble of the former on W3C, CERC and TU benchmarks. Significance of results is determined using a two-tailed paired randomization test \citep{Smucker2007} ($\HighlySignificant{} \, p < 0.01$; $\MoreSignificant{} \, p < 0.05$; $\Significant{} \, p < 0.1$) with respect to the ensemble ranking  (adjusted using the Benjamini-Hochberg procedure for multiple testing \citep{Benjamini1995}).\label{tbl:ensemble}}
	\resizebox{\textwidth}{!}{\input{resources/ensemble.tex}}
	\vspace*{-.5\baselineskip}
\end{table*}

The per-topic differences suggest that Model~2 and the log-linear model make very different errors: Model~2 excels at retrieving exact query matches, while the log-linear model performs semantic matching. Based on these observations we hypothesize that a combination of the two approaches will raise retrieval performance even further. To test this hypothesis, we propose a simple ensemble of rankings generated by Model~2 and the log-linear model by re-ranking candidates according to the multiplicatively-combined reciprocal rank:
\begin{equation}
\label{eq:ensemble}
\text{rank}_\text{ensemble}(c_j, q_i) \propto
 \frac{1}{\text{rank}_\text{model~2}(c_j, q_i)} \cdot \frac{1}{\text{rank}_\text{log-linear}(c_j, q_i)},
\end{equation}
where $\text{rank}_{M}(c_j, q_i)$ denotes the position of candidate $c_j$ in a ranking generated by model $M$ for answering query $q_i$. Equation~\eqref{eq:ensemble} is equivalent to performing data fusion using CombSUM \citep{Shaw1994} where the scores are given by the logarithm of the reciprocal ranks of the experts. Table~\ref{tbl:ensemble} compares the result of this ensemble to that of its constituents. Compared to the supervised methods of \citet{Fang2010}, we conclude that our fully unsupervised ensemble matches the performance of their method on the CERC 2007 benchmark and outperforms their method on the W3C 2005 benchmark. The superior performance of the ensemble suggests the viability of hybrid methods that combine semantic and exact matching.

\subsection{Scalability and efficiency}

Inference in the log-linear model is expressed in linear algebra operations (Section~\ref{section:methodology}). These operations can be efficiently performed by highly optimized software libraries and special-purpose hardware (i.e., GPUs). But the baseline methods against which we compare do not benefit from these speed-ups. Furthermore, many im\-plementation-specific details and choice of parameter values can influence runtime considerably (e.g. size of the latent representations). Therefore, we opt for a theoretical comparison of the inference complexity of the log-linear model and compare these to the baselines (Section~\ref{sec:baseline}).

The log-linear model generates a ranking of candidate experts by straight-forward matrix operations. The look-up operation in the projection matrix $\VocabularyMatrix{}$ occurs in constant time complexity, as the multiplication with the one-hot vector $v_i$ comes down to selecting the $i$-th column from $\VocabularyMatrix{}$. Multiplication of the $|C| \times \EmbeddingSize{}$ matrix $\CandidateMatrix{}$ with the $\EmbeddingSize{}$-dimensional word feature vector exhibits $O(|C| \cdot \EmbeddingSize{})$ runtime complexity. If we consider addition of the bias term and division by the normalizing function $Z_1$, the time complexity of \eqref{eq:unigram} becomes
\begin{equation*}
O(\underbrace{|C| \cdot (\EmbeddingSize{} + (\EmbeddingSize{} - 1))}_{\text{matrix-vector multiplication}} + \underbrace{|C|}_{\text{bias term}} + \underbrace{2 \cdot |C| - 1}_{Z_1}).
\end{equation*}
Notice, however, that the above analysis considers sequential execution. Modern computing hardware has the ability to parallelize common matrix operations \citep{Kruger2003,Fatahalian2004}. The number of candidate experts $|C|$ is the term that impacts performance most in the log-linear model (under the assumption that $|C| \gg \EmbeddingSize{}$).

If we consider $n$ terms, where $n$ is the query length during inference or the window size during training, then the complexity of~\eqref{eq:bag} becomes
\begin{eqnarray*}
\lefteqn{O(\underbrace{n \cdot |C| \cdot (2 \cdot \EmbeddingSize{} - 1) + n \cdot (3  \cdot |C| - 1)}_{\text{n forward-passes}}} \\
&&{} + \underbrace{(n - 1) \cdot |C|}_{\text{factor product}} + \underbrace{2 \cdot |C| - 1}_{Z_2})
\end{eqnarray*}
Notice that $Z_2$ does not need to be computed during inference as it does not affect the candidate expert ranking.

In terms of space complexity, parameters $\VocabularyMatrix{}$, $\CandidateMatrix{}$ and $\CandidateBias{}$, in addition to the intermediate results, all require memory space proportional to their size. Considering \eqref{eq:bag} for a sequence of $k$ words and batches of $m$ instances, we require $O(m \cdot k \cdot |C|)$ floating point numbers for every forward-pass to fit in-memory. While such an upper bound seems reasonable by modern computing standards, it is a severely limiting factor when considering large-scale communities and while utilizing limited-memory GPUs for fast computation.

The inferential complexity of the vector space-based models for entity retrieval \citep{Demartini2009} depends mainly on the dimensionality of the vectors and the number of candidate experts. The dimensionality of the latent entity representations is too high for efficient nearest neighbor retrieval \citep{Indyk1998} due to the curse of dimensionality. Therefore, the time complexity for the LSI- and TF-IDF-based vector space models are respectively $O(\gamma \cdot |C|)$ and $O(|V| \cdot |C|)$, where $\gamma$ denotes the number of latent topics in the LSI-based method. As hyperparameters $e$ and $\gamma$ both indicate the dimensionality of latent entity representations, the time complexity of the LSI-based method is comparable to that of the log-linear model. We note that $|V| \gg |C|$ for all benchmarks ($|V|$ is between $18$ to $91$ times larger than $|C|$) we consider in this paper and therefore conclude that the TF-IDF method loses to the log-linear model in terms of efficiency.

Compared to the unsupervised generative models of \citeauthor{Balog2006}, we have the profile-centric Model~1 and the document-centric Mo\-del~2 with inference time complexity $O(n \cdot |C|)$ and $O(n \cdot |D|)$, respectively, with $|D| \gg |C|$. In the previous section we showed that the log-linear model always performs better than Model~1 and nearly always outperforms Model~2. Hence, our log-linear model generally achieves the expertise retrieval performance of Model~2 (or higher) at the complexity cost of Model~1 during inference.

\subsection{Incremental indexing}

Existing unsupervised methods use well-understood maximum-likelihood language models that support incremental indexing. We now briefly discuss the incremental indexing capabilities of our proposed method. Extending the set of candidate experts $C$ requires the log-linear model to be re-trained from scratch as it changes the topology of the network. Moreover, every document associated with a candidate expert is considered as a negative example for all other candidates. While it is possible to reiterate over all past documents and only learn an additional row in matrix $\CandidateMatrix{}$, the final outcome is unpredictable.

If we consider a stream of documents instead of a predefined set $D$, the log-linear model can be learned in an online fashion. However, stochastic gradient descent requires that training examples are picked at random such that the batched update rule \eqref{eq:update} behaves like the empirical expectation over the full training set \citep{Bottou2010}. While we might be able to justify the assumption that documents arrive randomly, the $n$-grams extracted from those documents clearly violate this requirement.

Considering a stream of documents leads to the model forgetting expertise evidence as an (artificial) shift in the underlying distribution of the training data occurs. While such behavior is undesirable for the task considered in this paper, it might be well-suited for temporal expert finding \citep{Rybak2014,Fang2014}, where expertise drift over time is considered. However, temporal expertise finding is beyond the scope for this paper and left for future work.

%% file: resources/all_results.tex
\begin{tabular}{c@{ }c@{ }c@{ }c@{ }c@{ }c@{ }c@{ }c@{ }c@{ }c@{ }c}
\toprule
\multirow{2}{*}{W3C} & \multicolumn{5}{c}{2005} & \multicolumn{5}{c}{2006} \\ 
& MAP & NDCG@100 & MRR & P@5 & P@10 & MAP & NDCG@100 & MRR & P@5 & P@10 \\ 
\cmidrule(lr){2-6}
\cmidrule(lr){7-11}
\multicolumn{1}{l}{LSI} & $\phantom{}0.135$ & $\phantom{}0.266$ & $\phantom{}0.306$ & $\phantom{}0.192$ & $\phantom{}0.196$ & $\phantom{}0.245$ & $\phantom{}0.371$ & $\phantom{}0.482$ & $\phantom{}0.287$ & $\phantom{}0.338$ \\ 
\multicolumn{1}{l}{TF-IDF} & $\phantom{}0.243$ & $\phantom{}0.426$ & $\phantom{}0.541$ & $\phantom{}0.384$ & $\phantom{}0.350$ & $\phantom{}0.343$ & $\phantom{}0.531$ & $\phantom{}0.650$ & $\phantom{}0.492$ & $\phantom{}0.498$ \\ 
\multicolumn{1}{l}{Model 1 (d)} & $\phantom{}0.192$ & $\phantom{}0.358$ & $\phantom{}0.433$ & $\phantom{}0.276$ & $\phantom{}0.266$ & $\phantom{}0.321$ & $\phantom{}0.491$ & $\phantom{}0.635$ & $\phantom{}0.478$ & $\phantom{}0.449$ \\ 
\multicolumn{1}{l}{Model 1 (jm)} & $\phantom{}0.190$ & $\phantom{}0.352$ & $\phantom{}0.390$ & $\phantom{}0.272$ & $\phantom{}0.276$ & $\phantom{}0.311$ & $\phantom{}0.483$ & $\phantom{}0.596$ & $\phantom{}0.502$ & $\phantom{}0.437$ \\ 
\multicolumn{1}{l}{Model 2 (d)} & $\phantom{}0.198$ & $\phantom{}0.369$ & $\phantom{}0.429$ & $\phantom{}0.288$ & $\phantom{}0.272$ & $\phantom{}0.261$ & $\phantom{}0.419$ & $\phantom{}0.551$ & $\phantom{}0.441$ & $\phantom{}0.404$ \\ 
\multicolumn{1}{l}{Model 2 (jm)} & $\phantom{}0.211$ & $\phantom{}0.380$ & $\phantom{}0.451$ & $\phantom{}0.332$ & $\phantom{}0.296$ & $\phantom{}0.260$ & $\phantom{}0.423$ & $\phantom{}0.599$ & $\phantom{}0.449$ & $\phantom{}0.429$ \\ 
\multicolumn{1}{l}{Log-linear (ours)} & $\phantom{}\textbf{0.248}$ & $\phantom{}\textbf{0.444}$ & $\phantom{\Significant{}}\textbf{0.618}\Significant{}$ & $\phantom{}\textbf{0.412}$ & $\phantom{}\textbf{0.361}$ & $\phantom{\HighlySignificant{}}\textbf{0.484}\HighlySignificant{}$ & $\phantom{\HighlySignificant{}}\textbf{0.667}\HighlySignificant{}$ & $\phantom{\HighlySignificant{}}\textbf{0.833}\HighlySignificant{}$ & $\phantom{\HighlySignificant{}}\textbf{0.713}\HighlySignificant{}$ & $\phantom{\HighlySignificant{}}\textbf{0.644}\HighlySignificant{}$ \\ 
\cmidrule{1-11}
\multirow{2}{*}{CERC} & \multicolumn{5}{c}{2007} & \multicolumn{5}{c}{2008} \\ 
& MAP & NDCG@100 & MRR & P@5 & P@10 & MAP & NDCG@100 & MRR & P@5 & P@10 \\ 
\cmidrule(lr){2-6}
\cmidrule(lr){7-11}
\multicolumn{1}{l}{LSI} & $\phantom{}0.031$ & $\phantom{}0.107$ & $\phantom{}0.060$ & $\phantom{}0.016$ & $\phantom{}0.014$ & $\phantom{}0.038$ & $\phantom{}0.099$ & $\phantom{}0.106$ & $\phantom{}0.042$ & $\phantom{}0.055$ \\ 
\multicolumn{1}{l}{TF-IDF} & $\phantom{}0.332$ & $\phantom{}0.486$ & $\phantom{}0.463$ & $\phantom{}0.196$ & $\phantom{}0.141$ & $\phantom{}0.269$ & $\phantom{}0.465$ & $\phantom{}0.525$ & $\phantom{}0.332$ & $\phantom{}0.277$ \\ 
\multicolumn{1}{l}{Model 1 (d)} & $\phantom{}0.287$ & $\phantom{}0.427$ & $\phantom{}0.384$ & $\phantom{}0.156$ & $\phantom{}0.096$ & $\phantom{}0.181$ & $\phantom{}0.355$ & $\phantom{}0.388$ & $\phantom{}0.200$ & $\phantom{}0.172$ \\ 
\multicolumn{1}{l}{Model 1 (jm)} & $\phantom{}0.278$ & $\phantom{}0.420$ & $\phantom{}0.384$ & $\phantom{}0.156$ & $\phantom{}0.084$ & $\phantom{}0.170$ & $\phantom{}0.347$ & $\phantom{}0.339$ & $\phantom{}0.181$ & $\phantom{}0.159$ \\ 
\multicolumn{1}{l}{Model 2 (d)} & $\phantom{}0.352$ & $\phantom{}0.495$ & $\phantom{}0.454$ & $\phantom{}0.180$ & $\phantom{}0.138$ & $\phantom{}0.264$ & $\phantom{}0.461$ & $\phantom{}0.510$ & $\phantom{}0.281$ & $\phantom{}0.244$ \\ 
\multicolumn{1}{l}{Model 2 (jm)} & $\phantom{}\textbf{0.361}$ & $\phantom{}\textbf{0.500}$ & $\phantom{}0.467$ & $\phantom{}0.192$ & $\phantom{}0.138$ & $\phantom{}0.274$ & $\phantom{}0.463$ & $\phantom{}0.517$ & $\phantom{}0.278$ & $\phantom{}0.239$ \\ 
\multicolumn{1}{l}{Log-linear (ours)} & $\phantom{}0.344$ & $\phantom{}0.493$ & $\phantom{}\textbf{0.513}$ & $\phantom{}\textbf{0.215}$ & $\phantom{}\textbf{0.150}$ & $\phantom{\HighlySignificant{}}\textbf{0.342}\HighlySignificant{}$ & $\phantom{\MoreSignificant{}}\textbf{0.519}\MoreSignificant{}$ & $\phantom{\MoreSignificant{}}\textbf{0.656}\MoreSignificant{}$ & $\phantom{\Significant{}}\textbf{0.381}\Significant{}$ & $\phantom{}\textbf{0.299}$ \\ 
\cmidrule{1-11}
\multirow{2}{*}{TU} & \multicolumn{5}{c}{GT1} & \multicolumn{5}{c}{GT5} \\ 
& MAP & NDCG@100 & MRR & P@5 & P@10 & MAP & NDCG@100 & MRR & P@5 & P@10 \\ 
\cmidrule(lr){2-6}
\cmidrule(lr){7-11}
\multicolumn{1}{l}{LSI} & $\phantom{}0.095$ & $\phantom{}0.205$ & $\phantom{}0.153$ & $\phantom{}0.060$ & $\phantom{}0.051$ & $\phantom{}0.097$ & $\phantom{}0.208$ & $\phantom{}0.129$ & $\phantom{}0.043$ & $\phantom{}0.036$ \\ 
\multicolumn{1}{l}{TF-IDF} & $\phantom{}0.216$ & $\phantom{}0.356$ & $\phantom{}0.324$ & $\phantom{}0.131$ & $\phantom{}0.097$ & $\phantom{}0.233$ & $\phantom{}0.378$ & $\phantom{}0.288$ & $\phantom{}0.108$ & $\phantom{}0.079$ \\ 
\multicolumn{1}{l}{Model 1 (d)} & $\phantom{}0.171$ & $\phantom{}0.308$ & $\phantom{}0.258$ & $\phantom{}0.103$ & $\phantom{}0.082$ & $\phantom{}0.241$ & $\phantom{}0.385$ & $\phantom{}0.292$ & $\phantom{}0.109$ & $\phantom{}0.081$ \\ 
\multicolumn{1}{l}{Model 1 (jm)} & $\phantom{}0.189$ & $\phantom{}0.325$ & $\phantom{}0.277$ & $\phantom{}0.112$ & $\phantom{}0.085$ & $\phantom{}0.231$ & $\phantom{}0.373$ & $\phantom{}0.271$ & $\phantom{}0.100$ & $\phantom{}0.075$ \\ 
\multicolumn{1}{l}{Model 2 (d)} & $\phantom{}0.154$ & $\phantom{}0.284$ & $\phantom{}0.228$ & $\phantom{}0.087$ & $\phantom{}0.070$ & $\phantom{}0.191$ & $\phantom{}0.334$ & $\phantom{}0.233$ & $\phantom{}0.084$ & $\phantom{}0.065$ \\ 
\multicolumn{1}{l}{Model 2 (jm)} & $\phantom{}\textbf{0.234}$ & $\phantom{}\textbf{0.370}$ & $\phantom{}0.342$ & $\phantom{}0.136$ & $\phantom{}0.101$ & $\phantom{}0.253$ & $\phantom{}0.394$ & $\phantom{}0.302$ & $\phantom{}0.108$ & $\phantom{}0.081$ \\ 
\multicolumn{1}{l}{Log-linear (ours)} & $\phantom{}0.219$ & $\phantom{}0.356$ & $\phantom{}\textbf{0.351}$ & $\phantom{\Significant{}}\textbf{0.145}\Significant{}$ & $\phantom{}\textbf{0.105}$ & $\phantom{\HighlySignificant{}}\textbf{0.287}\HighlySignificant{}$ & $\phantom{\HighlySignificant{}}\textbf{0.425}\HighlySignificant{}$ & $\phantom{\HighlySignificant{}}\textbf{0.363}\HighlySignificant{}$ & $\phantom{\HighlySignificant{}}\textbf{0.134}\HighlySignificant{}$ & $\phantom{\HighlySignificant{}}\textbf{0.092}\HighlySignificant{}$ \\ 
\bottomrule
\end{tabular}

%% file: resources/ensemble.tex
\begin{tabular}{c@{ }c@{ }c@{ }c@{ }c@{ }c@{ }c@{ }c@{ }c@{ }c@{ }c}
\toprule
\multirow{2}{*}{W3C} & \multicolumn{5}{c}{2005} & \multicolumn{5}{c}{2006} \\ 
& MAP & NDCG@100 & MRR & P@5 & P@10 & MAP & NDCG@100 & MRR & P@5 & P@10 \\ 
\cmidrule(lr){2-6}
\cmidrule(lr){7-11}
\multicolumn{1}{l}{Model 2 (jm)} & $\phantom{}0.211$ & $\phantom{}0.380$ & $\phantom{}0.451$ & $\phantom{}0.332$ & $\phantom{}0.296$ & $\phantom{}0.260$ & $\phantom{}0.423$ & $\phantom{}0.599$ & $\phantom{}0.449$ & $\phantom{}0.429$ \\ 
\multicolumn{1}{l}{Log-linear (ours)} & $\phantom{}0.248$ & $\phantom{}0.444$ & $\phantom{}0.618$ & $\phantom{}0.412$ & $\phantom{}0.361$ & $\phantom{\HighlySignificant{}}\textbf{0.484}\HighlySignificant{}$ & $\phantom{\MoreSignificant{}}\textbf{0.667}\MoreSignificant{}$ & $\phantom{}\textbf{0.833}$ & $\phantom{\MoreSignificant{}}\textbf{0.713}\MoreSignificant{}$ & $\phantom{\MoreSignificant{}}\textbf{0.644}\MoreSignificant{}$ \\ 
\multicolumn{1}{l}{Ensemble} & $\phantom{\HighlySignificant{}}\textbf{0.291}\HighlySignificant{}$ & $\phantom{\MoreSignificant{}}\textbf{0.479}\MoreSignificant{}$ & $\phantom{}\textbf{0.668}$ & $\phantom{}\textbf{0.440}$ & $\phantom{}\textbf{0.378}$ & $\phantom{}0.433$ & $\phantom{}0.634$ & $\phantom{}0.825$ & $\phantom{}0.657$ & $\phantom{}0.586$ \\ 
\cmidrule{1-11}
\multirow{2}{*}{CERC} & \multicolumn{5}{c}{2007} & \multicolumn{5}{c}{2008} \\ 
& MAP & NDCG@100 & MRR & P@5 & P@10 & MAP & NDCG@100 & MRR & P@5 & P@10 \\ 
\cmidrule(lr){2-6}
\cmidrule(lr){7-11}
\multicolumn{1}{l}{Model 2 (jm)} & $\phantom{}0.361$ & $\phantom{}0.500$ & $\phantom{}0.467$ & $\phantom{}0.192$ & $\phantom{}0.138$ & $\phantom{}0.274$ & $\phantom{}0.463$ & $\phantom{}0.517$ & $\phantom{}0.278$ & $\phantom{}0.239$ \\ 
\multicolumn{1}{l}{Log-linear (ours)} & $\phantom{}0.344$ & $\phantom{}0.493$ & $\phantom{}0.513$ & $\phantom{}0.215$ & $\phantom{}0.150$ & $\phantom{}0.342$ & $\phantom{}0.519$ & $\phantom{}0.656$ & $\phantom{}0.381$ & $\phantom{}0.299$ \\ 
\multicolumn{1}{l}{Ensemble} & $\phantom{\MoreSignificant{}}\textbf{0.452}\MoreSignificant{}$ & $\phantom{\HighlySignificant{}}\textbf{0.589}\HighlySignificant{}$ & $\phantom{\HighlySignificant{}}\textbf{0.627}\HighlySignificant{}$ & $\phantom{\Significant{}}\textbf{0.248}\Significant{}$ & $\phantom{}\textbf{0.160}$ & $\phantom{\HighlySignificant{}}\textbf{0.395}\HighlySignificant{}$ & $\phantom{\HighlySignificant{}}\textbf{0.593}\HighlySignificant{}$ & $\phantom{}\textbf{0.716}$ & $\phantom{\MoreSignificant{}}\textbf{0.459}\MoreSignificant{}$ & $\phantom{\HighlySignificant{}}\textbf{0.357}\HighlySignificant{}$ \\ 
\cmidrule{1-11}
\multirow{2}{*}{TU} & \multicolumn{5}{c}{GT1} & \multicolumn{5}{c}{GT5} \\ 
& MAP & NDCG@100 & MRR & P@5 & P@10 & MAP & NDCG@100 & MRR & P@5 & P@10 \\ 
\cmidrule(lr){2-6}
\cmidrule(lr){7-11}
\multicolumn{1}{l}{Model 2 (jm)} & $\phantom{}0.234$ & $\phantom{}0.370$ & $\phantom{}0.342$ & $\phantom{}0.136$ & $\phantom{}0.101$ & $\phantom{}0.253$ & $\phantom{}0.394$ & $\phantom{}0.302$ & $\phantom{}0.108$ & $\phantom{}0.081$ \\ 
\multicolumn{1}{l}{Log-linear (ours)} & $\phantom{}0.219$ & $\phantom{}0.356$ & $\phantom{}0.351$ & $\phantom{}0.145$ & $\phantom{}0.105$ & $\phantom{}0.287$ & $\phantom{}0.425$ & $\phantom{}0.363$ & $\phantom{}0.134$ & $\phantom{}0.092$ \\ 
\multicolumn{1}{l}{Ensemble} & $\phantom{\HighlySignificant{}}\textbf{0.271}\HighlySignificant{}$ & $\phantom{\HighlySignificant{}}\textbf{0.417}\HighlySignificant{}$ & $\phantom{\HighlySignificant{}}\textbf{0.403}\HighlySignificant{}$ & $\phantom{\HighlySignificant{}}\textbf{0.165}\HighlySignificant{}$ & $\phantom{\HighlySignificant{}}\textbf{0.121}\HighlySignificant{}$ & $\phantom{\HighlySignificant{}}\textbf{0.331}\HighlySignificant{}$ & $\phantom{\HighlySignificant{}}\textbf{0.477}\HighlySignificant{}$ & $\phantom{\HighlySignificant{}}\textbf{0.402}\HighlySignificant{}$ & $\phantom{\HighlySignificant{}}\textbf{0.156}\HighlySignificant{}$ & $\phantom{\HighlySignificant{}}\textbf{0.105}\HighlySignificant{}$ \\ 
\bottomrule
\end{tabular}

%% file: conclusions.tex

\section{Conclusions}
\label{section:conclusions}

We have introduced an unsupervised discriminative, log-linear model for the expert retrieval task. Our approach exclusively employs raw textual evidence. Future work can focus on improving performance by feature engineering and incorporation of external evidence. Furthermore, no relevance feedback is required during training. This renders the model suitable for a broad range of applications and domains.

We evaluated our model on the W3C, CERC and TU benchmarks and compared it to state-of-the-art vector space-based entity ranking (based on LSI and TF-IDF) and language modeling (profile-centric and document-centric) approaches. The log-linear model combines the ranking performance of the best maximum-likelihood language modeling approach (document-centric) with inference time complexity linear in the number of candidate experts. We observed a notable increase in precision over existing methods. Analysis of our model's output reveals a negative correlation between the per-query performance and ranking uncertainty: higher confidence (i.e., lower entropy) in the rankings produced by our approach often occurs together with higher rank quality.

An error analysis of the log-linear model and traditional language models shows that the two make very different errors. These errors are mainly due to the semantic gap between query intent and the raw textual evidence. Some benchmarks expect exact query matches, others are helped by our semantic matching. An ensemble of methods employing exact and semantic matching generally outperforms the individual methods. This observation calls for further research in the area of combining exact and semantic matching.

One current limitation of our work is its scalability with respect to the number of candidate experts. We have started investigating trade-offs between model performance and time/space complexity. In the future we hope to apply scalable variants of this method on large-scale social media communities, for the purpose of determining topic ownership. While in this work we focus on expertise retrieval, the ideas we proposed can easily be transferred to the more general entity retrieval task. Moreover, our approach is likely to be applicable to authorship attribution and various other entity retrieval and prediction tasks.

%

\vspace*{\fill}

\begin{spacing}{-1}
\noindent
\small{
\textbf{Acknowledgments.}
We thank Isaac Sijaranamual, Manos Tsagkias, Tom Kenter, Zhaochun Ren, Ke Tran and Katja Hoffman for their useful comments and insights.

This research was supported by
Amsterdam Data Science,
the Dutch national program COMMIT,
Elsevier,
the European Community's Seventh Framework Programme (FP7/2007-2013) under
grant agreement nr 312827 (VOX-Pol),
the ESF Research Network Program ELIAS,
the Royal Dutch Academy of Sciences (KNAW) under the Elite Network Shifts project,
the Microsoft Research Ph.D.\ program,
the Netherlands eScience Center under project number 027.012.105,
the Netherlands Institute for Sound and Vision,
the Netherlands Organisation for Scientific Research (NWO)
under pro\-ject nrs
727.\-011.\-005, 
612.001.116, 
HOR-11-10, 
640.006.013, 
612.\-066.\-930, 
CI-14-25, 
SH-322-15, 
652.\-002.\-001, 
612.\-001.\-551, 
the Yahoo Faculty Research and Engagement Program,
and
Yandex.
All content represents the opinion of the authors, which is not necessarily shared or endorsed by their respective employers and/or sponsors.

Computing resources were provided by the Netherlands Organisation for Scientific Research (NWO) through allocation SH-322-15 of the Cartesius system and by the Advanced School for Computing and Imaging (ASCII) by allocation of the Distributed ASCII Supercomputer 4 (DAS-4) system.
}
\end{spacing}

%% file: appendix.tex

\newpage
\appendix

{

\newcommand{\BigSkip}{\qquad\qquad\qquad\qquad\qquad\qquad}

\newcommand{\LossGradNoReg}[1]{\frac{\partial \LossFn{}}{\partial #1} & = & - \frac{1}{m}  \Bigg( \sum_{i = 1}^m \frac{|\Document{}_{\max}|}{|\SourceDocument{i}|} \sum_{j = 1}^{|C|} P(\CandidateVector[j] | \SourceDocument{i}) \frac{\partial \log\left(P(\CandidateVector[j] \mid \Context[i])\right)}{\partial #1} \Bigg)}

\newcommand{\RescaleEqn}[1]{
	\vspace*{0.25\baselineskip}
	\scalebox{0.65}{
	\begin{minipage}{\linewidth}
	#1
	\end{minipage}
	}
	\vspace*{0.25\baselineskip}
}

The derivative of \eqref{eq:loss} w.r.t. bias term $\CandidateBias[]$ equals
\begin{center}
\RescaleEqn{
\begin{eqnarray*}
\hspace*{-1cm}
\LossGradNoReg{\CandidateBias[]} \nonumber
\end{eqnarray*}
}
\end{center}
and w.r.t. an arbitrary matrix parameter $\theta$ ($\VocabularyMatrix[]$ or $\CandidateMatrix[]$):
\begin{center}
\RescaleEqn{
\begin{eqnarray*}
\hspace*{-1cm}
\LossGradNoReg{\theta} \nonumber \\
& & + \frac{\lambda}{m} \sum_{i,j} \theta_{i,j} \nonumber.
\end{eqnarray*}
}
\end{center}
Further differentiation for parameter $\theta$ ($\VocabularyMatrix[]$, $\CandidateMatrix[]$ or $\CandidateBias[]$):
\begin{center}
\RescaleEqn{
\begin{eqnarray*}
\frac{\partial \log\left(P(\CandidateVector[j] \mid w_{1}, \ldots, w_{n})\right)}{\partial \theta} \nonumber
& =& \frac{1}{P(\CandidateVector[j] \mid w_{1}, \ldots, w_{n})} \frac{\partial P(\CandidateVector[j] \mid w_{1}, \ldots, w_{n})}{\partial \theta} \\
\frac{\partial P(\CandidateVector[j] \mid w_{1}, \ldots, w_{n})}{\partial \theta} & = &\frac{\frac{\partial \tilde{P}(c_j \mid w_{1}, \ldots, w_{n})}{\partial \theta} Z_2 - \tilde{P}(c_j \mid w_{1}, \ldots, w_{n}) \frac{\partial Z_2}{\partial \theta}}{Z^2_2} \\
\frac{\partial Z_2}{\partial \theta} &=& \sum_k \frac{\partial \tilde{P}(c_k \mid w_{1}, \ldots, w_{n})}{\partial \theta} \\
\frac{\partial \tilde{P}(c_j \mid w_{1}, \ldots, w_{n})}{\partial \theta}
&=& \sum_k \frac{\partial P(\CandidateVector[j] \mid w_k)}{\partial \theta} \prod_{i \neq k} P(\CandidateVector[j] \mid w_i)
\end{eqnarray*}
}
\end{center}
For a given candidate $c_j$ and word $w_i$, following \eqref{eq:unigram} we have
\newcommand{\NormalizedUnigramProb}{P(\CandidateVector[j] \mid w_i)}
\newcommand{\UnigramProb}{\tilde{P}(\CandidateVector[j] \mid w_i)}
\begin{center}
\RescaleEqn{
\begin{eqnarray*}
\NormalizedUnigramProb &=& \frac{\UnigramProb{}}{Z_1}\\
 &=& \frac{\exp \left( \left( \sum^{\EmbeddingSize{}}_{k=1} \CandidateMatrix[j,k] \VocabularyMatrix[k,i] \right) + \CandidateBias[j] \right)}{\sum^{|C|}_{l=1} \exp \left( \left( \sum^{\EmbeddingSize{}}_{k=1} \CandidateMatrix[l,k] \VocabularyMatrix[k,i] \right) + \CandidateBias[l] \right)}
\end{eqnarray*}
}
\end{center}
and consequently, with $\boldsymbol{\VocabularyMatrix[i]^\top}$ denoting the $i$-th column of matrix $\VocabularyMatrix[]$,
\begin{center}
\RescaleEqn{
\begin{equation}
\label{eq:dist_repr_grad}
\begin{split}
\frac{\partial \NormalizedUnigramProb}{\partial \boldsymbol{\CandidateMatrix[j]}} & = \frac{\left( Z_1 - \UnigramProb{} \right) \UnigramProb{} \boldsymbol{\VocabularyMatrix[i]^\top} }{Z^2_1} \\
\frac{\partial \NormalizedUnigramProb}{\partial \CandidateBias[j]} & = \frac{\left( Z_1 - \UnigramProb{} \right) \UnigramProb{}}{Z^2_1} \\
\frac{\partial \NormalizedUnigramProb}{\partial \boldsymbol{\VocabularyMatrix[i]^\top}} & = \frac{\left( \boldsymbol{\CandidateMatrix[j]} - \sum^{|C|}_{l=1} \boldsymbol{\CandidateMatrix[l]} \right) \UnigramProb{}}{Z_1}
\end{split}
\end{equation}
}
\end{center}
As can be seen in \eqref{eq:dist_repr_grad}, the distributed representations of candidates $c_j$ at time $t + 1$ are updated using the representation of words $w_i$ at time $t$ and vice versa.

}

%% file: arxiv16-expert-finding.bbl
\begin{thebibliography}{64}
\providecommand{\natexlab}[1]{#1}
\providecommand{\url}[1]{\texttt{#1}}
\expandafter\ifx\csname urlstyle\endcsname\relax
  \providecommand{\doi}[1]{doi: #1}\else
  \providecommand{\doi}{doi: \begingroup \urlstyle{rm}\Url}\fi

\bibitem[OEC(1996)]{OECD1996}
The knowledge-based economy.
\newblock Techn.\ report, Organisation for Economic Co-operation and
  Development, 1996.

\bibitem[Bailey et~al.(2007)Bailey, De~Vries, Craswell, and
  Soboroff]{Bailey2007}
P.~Bailey, A.~P. De~Vries, N.~Craswell, and I.~Soboroff.
\newblock Overview of the {TREC} 2007 enterprise track.
\newblock In \emph{TREC}, 2007.

\bibitem[Balog(2008)]{Balog2008}
K.~Balog.
\newblock \emph{People Search in the Enterprise}.
\newblock PhD thesis, University of Amsterdam, 2008.

\bibitem[Balog et~al.(2006)Balog, Azzopardi, and de~Rijke]{Balog2006}
K.~Balog, L.~Azzopardi, and M.~de~Rijke.
\newblock {Formal models for expert finding in enterprise corpora}.
\newblock In \emph{SIGIR}, pages 43--50, 2006.

\bibitem[Balog et~al.(2009)Balog, Azzopardi, and de~Rijke]{Balog2009}
K.~Balog, L.~Azzopardi, and M.~de~Rijke.
\newblock {A language modeling framework for expert finding}.
\newblock \emph{IPM}, 45:\penalty0 1--19, 2009.

\bibitem[Balog et~al.(2012)Balog, Fang, de~Rijke, Serdyukov, and Si]{Balog2012}
K.~Balog, Y.~Fang, M.~de~Rijke, P.~Serdyukov, and L.~Si.
\newblock Expertise retrieval.
\newblock \emph{Found. \& Tr. in Information Retrieval}, 6\penalty0
  (2-3):\penalty0 127--256, 2012.

\bibitem[Becerra-Fernandez(2000)]{BecerraFernandez2000}
I.~Becerra-Fernandez.
\newblock {Role of artificial intelligence technologies in the implementation
  of People-Finder knowledge management systems}.
\newblock \emph{Knowledge-Based Systems}, 13\penalty0 (5):\penalty0 315--320,
  2000.

\bibitem[Bengio et~al.(2003)Bengio, Ducharme, Vincent, and Janvin]{Bengio2003}
Y.~Bengio, R.~Ducharme, P.~Vincent, and C.~Janvin.
\newblock A neural probabilistic language model.
\newblock \emph{JMLR}, 3:\penalty0 1137--1155, 2003.

\bibitem[Benjamini and Hochberg(1995)]{Benjamini1995}
Y.~Benjamini and Y.~Hochberg.
\newblock Controlling the false discovery rate: a practical and powerful
  approach to multiple testing.
\newblock \emph{JSTOR}, pages 289--300, 1995.

\bibitem[Berendsen et~al.(2013)Berendsen, de~Rijke, Balog, Bogers, and van~den
  Bosch]{Berendsen2013}
R.~Berendsen, M.~de~Rijke, K.~Balog, T.~Bogers, and A.~van~den Bosch.
\newblock On the assessment of expertise profiles.
\newblock \emph{JASIST}, 64\penalty0 (10):\penalty0 2024--2044, 2013.

\bibitem[Bottou(2010)]{Bottou2010}
L.~Bottou.
\newblock Large-scale machine learning with stochastic gradient descent.
\newblock In \emph{COMPSTAT}, pages 177--186. Springer, 2010.

\bibitem[Burges et~al.(2005)Burges, Shaked, Renshaw, Lazier, Deeds, Hamilton,
  and Hullender]{Burges2005}
C.~Burges, T.~Shaked, E.~Renshaw, A.~Lazier, M.~Deeds, N.~Hamilton, and
  G.~Hullender.
\newblock {Learning to rank using gradient descent}.
\newblock In \emph{ICML}, pages 89--96, 2005.

\bibitem[Cao et~al.(2005)Cao, Liu, Bao, and Li]{Cao2005}
Y.~Cao, J.~Liu, S.~Bao, and H.~Li.
\newblock {Research on Expert Search at Enterprise Track of TREC 2005}.
\newblock In \emph{TREC}, pages 2--5, 2005.

\bibitem[Collobert et~al.(2011)Collobert, Weston, Bottou, Karlen, Kavukcuoglu,
  and Kuksa]{Collobert2011}
R.~Collobert, J.~Weston, L.~Bottou, M.~Karlen, K.~Kavukcuoglu, and P.~Kuksa.
\newblock Natural language processing (almost) from scratch.
\newblock \emph{JMLR}, 12\penalty0 (Aug):\penalty0 2493--2537, 2011.

\bibitem[Craswell et~al.(2001)Craswell, Hawking, Vercoustre, and
  Wilkins]{Craswell2001}
N.~Craswell, D.~Hawking, A.-M. Vercoustre, and P.~Wilkins.
\newblock P@noptic expert: {Searching} for experts not just for documents.
\newblock In \emph{Ausweb Poster Proceedings}, pages 21--25, 2001.

\bibitem[Craswell et~al.(2005)Craswell, de~Vries, and Soboroff]{Craswell2005}
N.~Craswell, A.~P. de~Vries, and I.~Soboroff.
\newblock Overview of the {TREC} 2005 enterprise track.
\newblock In \emph{TREC}, 2005.

\bibitem[Davenport and Prusak(1998)]{Davenport1998}
T.~H. Davenport and L.~Prusak.
\newblock \emph{Working Knowledge}.
\newblock Harvard Business Review Press, 1998.

\bibitem[Deerwester et~al.(1990)Deerwester, Dumais, and
  Harshman]{Deerwester1990}
S.~C. Deerwester, S.~T. Dumais, and R.~A. Harshman.
\newblock Indexing by latent semantic analysis.
\newblock \emph{JASIS}, 1990.

\bibitem[Demartini et~al.(2009)Demartini, Gaugaz, and Nejdl]{Demartini2009}
G.~Demartini, J.~Gaugaz, and W.~Nejdl.
\newblock A vector space model for ranking entities and its application to
  expert search.
\newblock In \emph{ECIR}, pages 189--201. Springer, 2009.

\bibitem[Deng et~al.(2013)Deng, He, and Gao]{Deng2013}
L.~Deng, X.~He, and J.~Gao.
\newblock {Deep stacking networks for information retrieval}.
\newblock In \emph{ICASSP}, pages 3153--3157, 2013.

\bibitem[Fang and Zhai(2007)]{Fang2007}
H.~Fang and C.~Zhai.
\newblock Probabilistic models for expert finding.
\newblock In \emph{ECIR}, pages 418--430, Berlin, Heidelberg, 2007.
  Springer-Verlag.

\bibitem[Fang and Godavarthy(2014)]{Fang2014}
Y.~Fang and A.~Godavarthy.
\newblock Modeling the dynamics of personal expertise.
\newblock In \emph{SIGIR}, pages 1107--1110, 2014.

\bibitem[Fang et~al.(2010)Fang, Si, and Mathur]{Fang2010}
Y.~Fang, L.~Si, and A.~P. Mathur.
\newblock {Discriminative models of integrating document evidence and
  document-candidate associations for expert search}.
\newblock In \emph{SIGIR}, pages 683--690, 2010.

\bibitem[Fatahalian et~al.(2004)Fatahalian, Sugerman, and
  Hanrahan]{Fatahalian2004}
K.~Fatahalian, J.~Sugerman, and P.~Hanrahan.
\newblock Understanding the efficiency of gpu algorithms for matrix-matrix
  multiplication.
\newblock In \emph{SIGGRAPH HWWS}, pages 133--137. ACM, 2004.

\bibitem[Glorot and Bengio(2010)]{Glorot2010}
X.~Glorot and Y.~Bengio.
\newblock Understanding the difficulty of training deep feedforward neural
  networks.
\newblock In \emph{AISTATS}, pages 249--256, 2010.

\bibitem[Hinton(1986)]{Hinton1986}
G.~E. Hinton.
\newblock Learning distributed representations of concepts.
\newblock In \emph{8th Annual Conference of the Cognitive Science Society},
  volume~1, page~12, Amherst, MA, 1986.

\bibitem[Hofmann(1999)]{Hofmann1999}
T.~Hofmann.
\newblock Probabilistic latent semantic indexing.
\newblock In \emph{SIGIR}, pages 50--57. ACM, 1999.

\bibitem[Huang et~al.(2013)Huang, Urbana, He, Gao, Deng, Acero, and
  Heck]{Huang2013}
P.-s. Huang, N.~M.~A. Urbana, X.~He, J.~Gao, L.~Deng, A.~Acero, and L.~Heck.
\newblock Learning deep structured semantic models for web search using
  clickthrough data.
\newblock In \emph{CIKM}, pages 2333--2338, 2013.

\bibitem[Indyk and Motwani(1998)]{Indyk1998}
P.~Indyk and R.~Motwani.
\newblock Approximate nearest neighbors: towards removing the curse of
  dimensionality.
\newblock In \emph{STOC}, pages 604--613. ACM, 1998.

\bibitem[Karimzadehgan and Zhai(2010)]{Karimzadehgan2010}
M.~Karimzadehgan and C.~Zhai.
\newblock Estimation of statistical translation models based on mutual
  information for ad hoc information retrieval.
\newblock In \emph{SIGIR}, pages 323--330. ACM, 2010.

\bibitem[Kiros et~al.(2014)Kiros, Salakhutdinov, and Zemel]{Kiros2014}
R.~Kiros, R.~Salakhutdinov, and R.~Zemel.
\newblock Multimodal neural language models.
\newblock In \emph{ICML}, pages 595--603, 2014.

\bibitem[Kruger and Dunning(1999)]{Kruger1991}
J.~Kruger and D.~Dunning.
\newblock Unskilled and unaware of it: how difficulties in recognizing one's
  own incompetence lead to inflated self-assessments.
\newblock \emph{J. Personality and Social Psych.}, 77\penalty0 (6):\penalty0
  1121, 1999.

\bibitem[Kr{\"u}ger and Westermann(2003)]{Kruger2003}
J.~Kr{\"u}ger and R.~Westermann.
\newblock Linear algebra operators for gpu implementation of numerical
  algorithms.
\newblock \emph{ACM Transactions on Graphics}, 22\penalty0 (3):\penalty0
  908--916, 2003.

\bibitem[Li and Xu(2014)]{Li2014}
H.~Li and J.~Xu.
\newblock Semantic matching in search.
\newblock \emph{Found. \& Tr. in Information Retrieval}, 7\penalty0
  (5):\penalty0 343--469, June 2014.

\bibitem[Liu(2011)]{Liu2011}
T.-Y. Liu.
\newblock \emph{Learning to Rank for Information Retrieval}.
\newblock Springer, 2011.

\bibitem[MacDonald and Ounis(2006)]{MacDonald2006}
C.~MacDonald and I.~Ounis.
\newblock {Voting for candidates: adapting data fusion techniques for an expert
  search task}.
\newblock In \emph{CIKM}, pages 387--396, 2006.

\bibitem[Macdonald and Ounis(2008)]{Macdonald2008}
C.~Macdonald and I.~Ounis.
\newblock {Expert search evaluation by supporting documents}.
\newblock In \emph{ECIR}, pages 555--563. Springer, 2008.

\bibitem[Maybury(2006)]{Maybury2006}
M.~T. Maybury.
\newblock Expert finding systems.
\newblock Techn.\ Report MTR-06B000040, MITRE, 2006.

\bibitem[McDonald and Ackerman(2000)]{McDonald2000}
D.~W. McDonald and M.~S. Ackerman.
\newblock {Expertise recommender}.
\newblock In \emph{CSCW}, pages 231--240, 2000.

\bibitem[Mikolov et~al.(2010)Mikolov, Karafiat, Burget, Cernocky, and
  Khudanpur]{Mikolov2010}
T.~Mikolov, M.~Karafiat, L.~Burget, J.~Cernocky, and S.~Khudanpur.
\newblock Recurrent neural network based language model.
\newblock In \emph{Interspeech}, pages 1045--1048, 2010.

\bibitem[Mikolov et~al.(2013{\natexlab{a}})Mikolov, Chen, Corrado, and
  Dean]{Mikolov2013-2}
T.~Mikolov, K.~Chen, G.~Corrado, and J.~Dean.
\newblock Distributed representations of words and phrases and their
  compositionality.
\newblock In \emph{NIPS}, pages 3111--3119, 2013{\natexlab{a}}.

\bibitem[Mikolov et~al.(2013{\natexlab{b}})Mikolov, Corrado, Chen, and
  Dean]{Mikolov2013}
T.~Mikolov, G.~Corrado, K.~Chen, and J.~Dean.
\newblock Efficient estimation of word representations in vector space.
\newblock arXiv 1301.3781, 2013{\natexlab{b}}.

\bibitem[Mnih and Hinton(2007)]{Mnih2007}
A.~Mnih and G.~Hinton.
\newblock {Three new graphical models for statistical language modelling}.
\newblock In \emph{ICML}, pages 641--648, 2007.

\bibitem[Mnih and Hinton(2008)]{Mnih2009}
A.~Mnih and G.~Hinton.
\newblock A scalable hierarchical distributed language model.
\newblock In \emph{NIPS}, pages 1081--1088, 2008.

\bibitem[Mnih and Kavukcuoglu(2013)]{Mnih2013}
A.~Mnih and K.~Kavukcuoglu.
\newblock {Learning word embeddings efficiently with noise-contrastive
  estimation}.
\newblock In \emph{NIPS}, pages 2265--2273, 2013.

\bibitem[Montavon et~al.(2012)Montavon, Orr, and M{\"u}ller]{Montavon2012}
G.~Montavon, G.~B. Orr, and K.-R. M{\"u}ller.
\newblock \emph{Neural Networks: Tricks of the Trade}.
\newblock Springer, 2012.

\bibitem[Moreira et~al.(2011)Moreira, Martins, and Calado]{Moreira2011}
C.~Moreira, B.~Martins, and P.~Calado.
\newblock Using rank aggregation for expert search in academic digital
  libraries.
\newblock In \emph{Simp\'{o}sio de Inform\'{a}tica, INForum}, pages 1--10,
  2011.

\bibitem[Pennington et~al.(2014)Pennington, Socher, and
  Manning]{Pennington2014}
J.~Pennington, R.~Socher, and C.~D. Manning.
\newblock {GloVe: Global Vectors for Word Representation}.
\newblock In \emph{EMNLP}, pages 1532--1543, 2014.

\bibitem[Petkova and Croft(2006)]{Petkova2006}
D.~Petkova and W.~B. Croft.
\newblock {Hierarchical language models for expert finding in enterprise
  corpora}.
\newblock In \emph{ICTAI '06}, pages 599--606, 2006.

\bibitem[Powell and Snellman(2004)]{Powell2004}
W.~W. Powell and K.~Snellman.
\newblock The knowledge economy.
\newblock \emph{Annual review of sociology}, pages 199--220, 2004.

\bibitem[Rumelhart et~al.(1986)Rumelhart, Hinton, and Williams]{Rumelhart1986}
D.~Rumelhart, G.~Hinton, and R.~Williams.
\newblock {Learning internal representations by back propagation}.
\newblock In \emph{Parallel Distributed Processing}, pages 318--362. MIT Press,
  1986.

\bibitem[Rybak et~al.(2014)Rybak, Balog, and N{\o}rv{\aa}g]{Rybak2014}
J.~Rybak, K.~Balog, and K.~N{\o}rv{\aa}g.
\newblock Temporal expertise profiling.
\newblock In \emph{ECIR}, pages 540--546. Springer, 2014.

\bibitem[Salakhutdinov and Hinton(2009)]{Salakhutdinov2009}
R.~Salakhutdinov and G.~Hinton.
\newblock Semantic hashing.
\newblock \emph{Int. J. Approximate Reasoning}, 50\penalty0 (7):\penalty0
  969--978, 2009.

\bibitem[Serdyukov and Hiemstra(2008)]{Serdyukov2008}
P.~Serdyukov and D.~Hiemstra.
\newblock Modeling documents as mixtures of persons for expert finding.
\newblock In \emph{ECIR}, pages 309--320. Springer, 2008.

\bibitem[Serdyukov et~al.(2008)Serdyukov, Rode, and Hiemstra]{Serdyukov2008-2}
P.~Serdyukov, H.~Rode, and D.~Hiemstra.
\newblock {Modeling multi-step relevance propagation for expert finding}.
\newblock In \emph{CIKM}, pages 1133--1142, 2008.

\bibitem[Shannon(1948)]{Shannon1948}
C.~Shannon.
\newblock A mathematical theory of communication.
\newblock \emph{Bell System Technical J.}, 27:\penalty0 379--423, 623--656,
  1948.

\bibitem[Shaw et~al.(1994)Shaw, Fox, Shaw, and Fox]{Shaw1994}
J.~A. Shaw, E.~A. Fox, J.~A. Shaw, and E.~A. Fox.
\newblock Combination of multiple searches.
\newblock In \emph{TREC}, pages 243--252, 1994.

\bibitem[Shen et~al.(2014)Shen, He, Gao, Deng, and Mesnil]{Shen2014}
Y.~Shen, X.~He, J.~Gao, L.~Deng, and G.~Mesnil.
\newblock A latent semantic model with convolutional-pooling structure for
  information retrieval.
\newblock In \emph{CIKM}, pages 101--110, 2014.

\bibitem[Smucker et~al.(2007)Smucker, Allan, and Carterette]{Smucker2007}
M.~D. Smucker, J.~Allan, and B.~Carterette.
\newblock A comparison of statistical significance tests for information
  retrieval evaluation.
\newblock In \emph{CIKM}, pages 623--632. ACM, 2007.

\bibitem[Sorg and Cimiano(2011)]{Sorg2011}
P.~Sorg and P.~Cimiano.
\newblock {Finding the right expert: Discriminative models for expert
  retrieval}.
\newblock In \emph{KDIR}, pages 190--199, 2011.

\bibitem[TREC(2005--2008)]{TREC2010}
TREC.
\newblock {Enterprise Track}, 2005--2008.

\bibitem[van Dijk et~al.(2015)van Dijk, Tsagkias, and de~Rijke]{vanDijk2015}
D.~van Dijk, M.~Tsagkias, and M.~de~Rijke.
\newblock Early detection of topical expertise in community question and
  answering.
\newblock In \emph{SIGIR}, 2015.

\bibitem[Vapnik(1998)]{Vapnik1998}
V.~Vapnik.
\newblock \emph{Statistical learning theory}, volume~1.
\newblock Wiley New York, 1998.

\bibitem[Zeiler(2012)]{Zeiler2012}
M.~D. Zeiler.
\newblock Adadelta: An adaptive learning rate method.
\newblock \emph{CoRR}, abs/1212.5701, 2012.

\end{thebibliography}
